\documentclass[12pt]{iopart}

\usepackage{iopams} 
\usepackage{amsthm} 
\usepackage{epsf}

\eqnobysec

\newcommand{\Sw}{{\cal S}(\mathbb R \frac{\ }{\ } \{ a,b \} )}
\newcommand{\Swt}{{\cal S}^{\times}(\mathbb R \frac{\ }{\ } \{ a,b \} )}
\newcommand{\Swp}{{\cal S}^{\prime}(\mathbb R \frac{\ }{\ } \{ a,b \} )}
\newcommand{\rhsSwt}{\Sw \subset L^2 \subset \Swt}

\begin{document}

\def\llra{\relbar\joinrel\longrightarrow}              
\def\mapright#1{\smash{\mathop{\llra}\limits_{#1}}}    
\def\mapup#1{\smash{\mathop{\llra}\limits^{#1}}}     
\def\mapupdown#1#2{\smash{\mathop{\llra}\limits^{#1}_{#2}}} 

%
\catcode`\@=11

\def\BF#1{{\bf {#1}}}
\def\NEG#1{{\rlap/#1}}

\def\Let@{\relax\iffalse{\fi\let\\=\cr\iffalse}\fi}
\def\vspace@{\def\vspace##1{\crcr\noalign{\vskip##1\relax}}}
\def\multilimits@{\bgroup\vspace@\Let@
 \baselineskip\fontdimen10 \scriptfont\tw@
 \advance\baselineskip\fontdimen12 \scriptfont\tw@
 \lineskip\thr@@\fontdimen8 \scriptfont\thr@@
 \lineskiplimit\lineskip
 \vbox\bgroup\ialign\bgroup\hfil$\m@th\scriptstyle{##}$\hfil\crcr}
\def\Sb{_\multilimits@}
\def\endSb{\crcr\egroup\egroup\egroup}
\def\Sp{^\multilimits@}
\let\endSp\endSb
%

\title[The RHS in Quantum Mechanics]{The role of the rigged Hilbert space in 
Quantum Mechanics}

\author{Rafael de la Madrid}
\address{Departamento de F\'\i sica Te\'orica, Facultad de Ciencias,
Universidad del Pa\'\i s Vasco, 48080 Bilbao, Spain \\
E-mail: {\texttt{wtbdemor@lg.ehu.es}}}

\date{\small{January 4, 2005}}

\begin{abstract}
There is compelling evidence that, when continuous spectrum is 
present, the natural mathematical setting for Quantum Mechanics is the rigged 
Hilbert space rather than just the Hilbert space. In particular, Dirac's 
bra-ket formalism is fully implemented by the rigged Hilbert space rather than
just by the Hilbert space. In this paper, we provide a pedestrian introduction
to the role the rigged Hilbert space plays in Quantum Mechanics, by way of a 
simple, exactly solvable example. The procedure will be constructive and based 
on a recent publication. We also provide a thorough discussion 
on the physical significance of the rigged Hilbert space.
\end{abstract}


\maketitle

\section{Introduction}
\setcounter{equation}{0}
\label{sec:introduction}

It has been known for several decades that Dirac's bra-ket formalism is 
mathematically justified not by the Hilbert space alone, but by the
rigged Hilbert space (RHS). This is the reason why there is an increasing 
number of Quantum Mechanics textbooks that already include the 
rigged Hilbert space as part of their contents (see, for example, 
Refs.~\cite{ATKINSON}-\cite{KUKULIN}). Despite the importance of the RHS, 
there is still a lack of 
simple examples for which the corresponding RHS is constructed in a didactical 
manner. Even worse, there is no pedagogical
discussion on the physical significance of the RHS. In this paper, we
use the one-dimensional (1D) rectangular barrier potential to introduce the RHS
at the graduate student level. As well, we discuss the physical significance 
of each of the ingredients that form the RHS. The construction of the RHS of 
such a simple model will unambiguously show that the RHS is needed at the 
most basic level of Quantum Mechanics.

The present paper is complemented by a previous publication, 
Ref.~\cite{FOCO}, to which we shall refer the reader interested in a 
detailed mathematical account on the construction of the RHS of the 1D 
rectangular barrier. For a general background on the Hilbert and the rigged 
Hilbert space methods, the reader may consult Ref.~\cite{DIS} and 
references therein.

Dirac's bra-ket formalism was introduced by Dirac in his classic
monograph~\cite{DIRAC}. Since its inception, Dirac's abstract algebraic model 
of {\it bras} and {\it kets} (from the bracket notation for the inner 
product) proved to be of great calculational value, although there were 
serious difficulties in finding a mathematical justification for the actual 
calculations within the Hilbert space, as Dirac~\cite{DIRAC} and von 
Neumann~\cite{VON} themselves state in their books~\cite{QUOTEVONDIRAC}. As 
part of his bra-ket formalism, Dirac introduced the so-called 
Dirac delta function, a formal entity without a counterpart in the classical
theory of functions. It was L.~Schwartz who gave a precise meaning to the 
Dirac delta function as a functional over a space of test 
functions~\cite{SCHWARTZ}. This led to the development of a new branch of 
functional analysis, the theory of distributions. By combining von Neumann's 
Hilbert space with the theory of distributions, I.~Gelfand and collaborators 
introduced the RHS~\cite{GELFAND,MAURIN}. It was already clear to the creators 
of the RHS that their formulation was the mathematical support of Dirac's 
bra-ket formalism~\cite{CITEMAURIN}. The RHS made its first appearance in the 
Physics literature in the 1960s~\cite{ROBERTS,ANTOINE,B60}, when some 
physicists also realized that the RHS provides a rigorous mathematical 
rephrasing of all of the aspects of Dirac's bra-ket formalism. Nowadays,
there is a growing consensus that the RHS, rather than the Hilbert space
alone, is the natural mathematical setting of Quantum 
Mechanics~\cite{QUOTEBALLENTINE}.

A note on semantics. The word ``rigged'' in rigged Hilbert space has a 
nautical connotation, such as the phrase ``fully rigged ship;'' it
has nothing to do with any unsavory practice such as ``fixing'' or 
predetermining a result. The phrase ``rigged Hilbert space'' is a direct
translation of the phrase ``osnashchyonnoe Hilbertovo prostranstvo'' from
the original Russian. A more faithful translation would be
``equipped Hilbert space.'' Indeed, the rigged Hilbert space is just the 
Hilbert space equipped with distribution theory---in Quantum Mechanics, to 
rig a Hilbert space means simply to equip that Hilbert space with distribution
theory. Thus, the RHS is not a replacement but an enlargement of the Hilbert 
space. 

The RHS is {\it neither} an extension {\it nor} an 
interpretation of the physical principles of Quantum Mechanics, but rather 
the most natural, concise and logic language to formulate Quantum 
Mechanics. The RHS is simply a mathematical tool to extract 
and process the information contained in observables 
that have continuous spectrum. Observables with discrete spectrum and a 
finite number of eigenvectors (e.g., spin) do not need the RHS. For such 
observables, the Hilbert space is sufficient. Actually, as we shall explain, 
in general only unbounded observables with continuous spectrum need the RHS.

The usefulness of the RHS is not simply restricted to accounting for Dirac's
bra-ket formalism. The RHS has also proved to be a very useful 
research tool in the quantum theory of scattering and decay 
(see Ref.~\cite{DIS} and references therein), and in the construction of 
generalized spectral decompositions of chaotic maps~\cite{AT93,SUCHANECKI}. In
fact, it seems that the RHS is the natural language to deal with problems
that involve continuous and resonance spectra. 

Loosely speaking, a rigged Hilbert space (also called a Gelfand triplet) is
a triad of spaces
\begin{equation}
       {\mathbf \Phi} \subset {\cal H} \subset {\mathbf \Phi}^{\times}
      \label{RHStIntro}
\end{equation}
such that $\cal H$ is a Hilbert space, $\mathbf \Phi$ is a dense
subspace of $\cal H$~\cite{DENSE}, and $\mathbf \Phi ^{\times}$ is the space of
antilinear functionals over $\mathbf \Phi$~\cite{FUNCTIONAL}. Mathematically,
$\mathbf \Phi$ is the space of test functions, and $\mathbf \Phi ^{\times}$
is the space of distributions. The space $\mathbf \Phi ^{\times}$ is called
the antidual space of $\mathbf \Phi$. Associated with the 
RHS~(\ref{RHStIntro}), there is always another RHS,
\begin{equation}
       {\mathbf \Phi} \subset {\cal H} \subset {\mathbf \Phi}^{\prime} \, ,
      \label{RHSpIntro}
\end{equation}
where ${\mathbf \Phi}^{\prime}$ is called the dual space of ${\mathbf \Phi}$
and contains the linear functionals over $\mathbf \Phi$~\cite{FUNCTIONAL}. 

The basic reason why we need the spaces ${\mathbf \Phi}^{\prime}$ and
${\mathbf \Phi}^{\times}$ is that the bras and kets associated with the 
elements in the continuous spectrum of an observable belong, respectively, to 
${\mathbf \Phi}^{\prime}$ and ${\mathbf \Phi}^{\times}$ rather than to
${\cal H}$. The basic reason reason why we need the space $\mathbf \Phi$ is 
that unbounded operators are not defined on the whole of ${\cal H}$ but only
on dense subdomains of ${\cal H}$ that are not invariant under the
action of the observables. Such non-invariance makes expectation values,
uncertainties and commutation relations not well defined on the whole
of $\cal H$. The space $\mathbf \Phi$ is the largest subspace of the Hilbert 
space on which such expectation values, uncertainties and commutation 
relations are well defined.

The original formulation of the RHS~\cite{GELFAND,MAURIN} does not provide a 
systematic procedure to construct the RHS generated by the Hamiltonian of the 
Schr\"odinger equation, since the space $\mathbf \Phi$ is assumed to be
given beforehand. Such systematic procedure is important because,
after all, claiming that the RHS is the natural setting for 
Quantum Mechanics is about the same as claiming that, when the Hamiltonian
has continuous spectrum, the natural setting for the solutions of the
Schr\"odinger equation is the RHS rather than just the Hilbert space. The 
task of developing a systematic procedure to construct the RHS generated 
by the Schr\"odinger equation was undertaken in Ref.~\cite{DIS}. The 
method proposed in Ref.~\cite{DIS}, which was partly based on 
Refs.~\cite{ROBERTS,ANTOINE,B60}, has been applied to two simple
three-dimensional potentials, see Refs.~\cite{JPA02,FP02}, to the
three-dimensional free Hamiltonian, see Ref.~\cite{IJTP03}, and to
the 1D rectangular barrier potential, see Ref.~\cite{FOCO}. In this paper, we 
present the method of Ref.~\cite{DIS} in a didactical manner.

The organization of the paper is as follows. In Sec.~\ref{sec:why}, we outline
the major reasons why the RHS provides the mathematical setting for Quantum
Mechanics. In Sec.~\ref{sec:e1dsbp}, we recall the basics of the
1D rectangular potential model. Section~\ref{sec:crhs} provides the RHS of
this model. In Sec.~\ref{sec:phymean}, we discuss the physical meaning of
each of the ingredients that form the RHS. In Sec.~\ref{sec:gener}, we 
discuss the relation of the Hilbert space spectral measures with the bras
and kets, as well as the limitations of our method to construct
RHSs. Finally, Sec.~\ref{sec:conclusions} contains the conclusions to the 
paper.

\section{Motivating the rigged Hilbert space}
\label{sec:why}

The {\it linear superposition principle} and the 
{\it probabilistic interpretation} of Quantum Mechanics are two 
major guiding principles in our 
understanding of the microscopic world. These two principles suggest that the
space of states be a linear space (which accounts for the superposition
principle) endowed with a scalar product (which is used to calculate 
probability amplitudes). A linear space endowed with a scalar product is 
called a Hilbert space and is usually denoted by $\cal H$~\cite{HSDEF}.

In Quantum Mechanics, observable quantities are represented
by linear, self-adjoint operators acting on $\cal H$. The eigenvalues 
of an operator represent the possible values of the 
measurement of the corresponding observable. These eigenvalues, which 
mathematically correspond to the spectrum of the operator, can be discrete 
(as the energies of a particle in a box), continuous (as the energies 
of a free, unconstrained particle), or a combination of
discrete and continuous (as the energies of the Hydrogen atom).

When the spectrum of an observable $A$ is discrete and $A$ is 
bounded~\cite{UNB}, then $A$ is defined on the whole of $\cal H$ and
the eigenvectors of $A$ belong to $\cal H$. In this case, $A$ can be 
essentially seen as a matrix. This means that, as 
far as discrete spectrum is concerned, there is no need to extend 
$\cal H$. However, quantum mechanical
observables are in general unbounded~\cite{UNB} and their spectrum has 
in general a continuous part. In order to deal with continuous 
spectrum, textbooks usually follow Dirac's bra-ket 
formalism, which is a heuristic generalization of the linear algebra of
Hermitian matrices used for discrete spectrum. As we shall see, the 
mathematical methods of the Hilbert space are not sufficient to make sense of 
the prescriptions of Dirac's formalism, the reason for which we shall 
extend the Hilbert space to the rigged Hilbert space.

For pedagogical reasons, we recall the essentials of the linear algebra of 
Hermitian matrices before proceeding with Dirac's formalism.

\subsection{Hermitian matrices}

If the measurement of an observable $A$ (e.g., spin) yields a discrete,
finite number $N$ of results $a_n$, $n=1, 2, \ldots , N$, then $A$ is realized 
by a Hermitian matrix on a Hilbert space $\cal H$ of dimension $N$. Since 
$\cal H$ is an $N$-dimensional linear 
space, there are $N$ linearly independent vectors $\{ e_n \} _{n=1}^N$ that 
form an orthonormal basis system for $\cal H$. We denote these basis vectors 
$e_n$ also by $|e_n\rangle$. The scalar products of the elements of the basis 
system are written in one of the following ways:
\begin{equation} 
       e_n \cdot e_m \equiv (e_n,e_m)\equiv \langle e_n|e_m\rangle =
       \delta_{nm} \, , \qquad n,m=1,2,\ldots , N \, ,
       \label{dcnII}
\end{equation} 
where $\delta_{nm}$ is the Kronecker delta. As the basis system for the space 
$\cal H$, it is always possible to choose the eigenvectors of 
$A$. Therefore, one can choose basis vectors $e_n\in {\cal H}$ which also 
fulfill
\begin{equation} 
       Ae_n=a_n e_n \, .
\end{equation} 
Since $A$ is Hermitian, the eigenvalues $a_n$ are real. The eigenvectors $e_n$ 
are often labeled by their eigenvalues $a_n$ and denoted by
\begin{equation} 
      e_n\equiv |a_n\rangle \, ,
\end{equation} 
and they are represented by column vectors. For each column eigenvector
$e_n\equiv |a_n\rangle$, there also exists a row eigenvector
$\tilde{e}_n\equiv \langle a_n|$ that is a left eigenvector of $A$,
\begin{equation} 
      \tilde{e}_n A = a_n \tilde{e}_n \, .
\end{equation} 
Thus, when $A$ is a Hermitian matrix acting on an $N$-dimensional 
Hilbert space $\mathcal H$, for each eigenvalue $a_n$ of $A$ there exist 
a right (i.e., column) eigenvector of $A$
\begin{equation}
     A|a_n\rangle =a_n|a_n \rangle \, , \quad n=1,2,\ldots,N \, , 
                \label{fddeII}  
\end{equation}
and also a left (i.e., row) eigenvector of $A$
\begin{equation}
     \langle a_n|A =a_n \langle a_n| \, , \quad n=1,2,\ldots,N \, , 
                \label{fddeIIl}  
\end{equation}
such that these row and column eigenvectors are orthonormal,
\begin{equation} 
      \langle a_n|a_m \rangle =\delta _{nm} \, , \quad n,m=1,2,\ldots ,N \, ,
      \label{orthonomr}
\end{equation}
and such that every vector $\varphi \in {\cal H}$ can be written as
\begin{equation}
      \varphi=\sum^N_{n=1}|a_n\rangle \langle a_n|\varphi \rangle \, . 
      \label{fddsdII}
\end{equation}
Equation~(\ref{fddsdII}) is called the eigenvector expansion 
of $\varphi$ with respect to the eigenvectors of $A$. The complex numbers
$\langle a_n|\varphi \rangle$ are the components of the vector $\varphi$ 
with respect to the basis of eigenvectors of $A$. Physically, 
$\langle a_n|\varphi \rangle$ represents the probability amplitude of 
obtaining the value $a_n$ in the measurement of the observable $A$ on the
state $\varphi$. By acting on both sides of Eq.~(\ref{fddsdII}) with $A$,
and recalling Eq.~(\ref{fddeII}), we obtain that
\begin{equation}
     A \varphi=\sum^N_{n=1}a_n |a_n\rangle \langle a_n|\varphi \rangle \, . 
      \label{fddsdIIA}
\end{equation}

\subsection{Dirac's bra-ket formalism}

Dirac's formalism is an elegant, heuristic generalization of the algebra of 
finite dimensional matrices to the continuous-spectrum, infinite-dimensional 
case. Four of the most important features of Dirac's formalism are:

\begin{enumerate}
      \item To each element of the spectrum of an observable $A$, 
there correspond a left and a right eigenvector (for the moment, we assume
that the spectrum is non-degenerate). If discrete eigenvalues are 
denoted by $a_n$ and continuous eigenvalues by $a$, then the
corresponding right eigenvectors, which are denoted by the kets 
$|a_n\rangle$ and $|a \rangle$, satisfy 
\numparts
    \begin{equation}
           A|a_n \rangle =a_n|a_n\rangle \, , 
             \label{dketqueintro}  
    \end{equation}
    \begin{equation}
           A|a \rangle =a |a \rangle \, ,
           \label{cketequeintro}
    \end{equation}
\endnumparts
and the corresponding left eigenvectors, which are denoted 
by the bras $\langle a_n|$ and $\langle a|$, satisfy
\numparts
      \begin{equation}
         \langle a_n|A=a_n \langle a_n| \, ,  
    \end{equation}
    \begin{equation}
         \langle a|A=a \langle a| \, .
          \label{braeqeneintro}
    \end{equation}
\endnumparts
The bras $\langle a|$ generalize the notion of row eigenvectors, whereas the 
kets $|a \rangle$ generalize the notion of column eigenvectors.

   \item In analogy to Eq.~(\ref{fddsdII}), the eigenbras and eigenkets of 
an observable form a complete basis, that is, any wave function $\varphi$ can 
be expanded in the so-called Dirac basis expansion:
    \begin{equation}
       \varphi = \sum_n |a_n\rangle \langle a_n|\varphi \rangle + 
       \int \rmd a \, |a \rangle \langle a |\varphi \rangle \, .
        \label{introDirbaexp}
      \end{equation}
In addition, the bras and kets furnish a resolution of the identity,
   \begin{equation}
       I = \sum_n |a_n\rangle \langle a_n| + 
       \int \rmd a \, |a \rangle \langle a | \, ,
        \label{introresiden}
      \end{equation}
and, in a generalization of Eq.~(\ref{fddsdIIA}), the action of $A$ can be 
written as
\begin{equation}
       A = \sum_n a_n |a_n\rangle \langle a_n| + 
       \int \rmd a \, a |a \rangle \langle a | \, .
        \label{introactionA}
      \end{equation}

 \item The bras and kets are normalized according to the following rule:
   \numparts      
         \begin{equation}
              \langle a_n|a_m\rangle =\delta _{nm} \, , 
          \end{equation}
         \begin{equation}
           \langle a |a ^{\prime} \rangle = \delta (a -a ^{\prime}) \, ,
            \label{deltanorintro}
         \end{equation}
   \endnumparts
where $\delta _{nm}$ is the Kronecker delta and $\delta (a-a ^{\prime})$ is 
the Dirac delta. The Dirac delta normalization generalizes the 
orthonormality~(\ref{orthonomr}) of the eigenvectors of a Hermitian matrix.
      \item  Like in the case of two finite-dimensional matrices,
all algebraic operations such as the commutator of two observables $A$ and $B$,
     \begin{equation}
           [A,B]=AB-BA \, ,
           \label{comuts}
     \end{equation}
are always well defined.
\end{enumerate}

\subsection{The need of the rigged Hilbert space}

In Quantum Mechanics, observables are usually given by differential 
operators. In the Hilbert space framework, the formal prescription of an 
observable leads to the definition of a linear operator as follows: One has 
to find 
first the Hilbert space $\cal H$, then one sees on what elements of $\cal H$ 
the action of the observable makes sense, and finally one checks whether the 
action of the observable remains in $\cal H$. For example, the position 
observable $Q$ of a 1D particle is given by
\begin{equation}
      Qf(x)=xf(x) \, .
        \label{fdopx}
\end{equation}
The Hilbert space of a 1D particle is given by the collection of square 
integrable functions,
\begin{equation}
     L^2 =\{ f(x) \, | \ \int_{-\infty}^{\infty}\rmd x \, 
                  |f(x)|^2 < \infty  \}   \, ,
     \label{l2space}
\end{equation}
and the action of $Q$, although in principle well defined on every
element of $L^2$, remains in $L^2$
only for the elements of the following subspace:
\begin{equation}
      {\cal D}(Q)= \{ f(x) \in L^2   \, | \ 
            \int_{-\infty}^{\infty}\rmd x \, 
                  |xf(x)|^2 < \infty  \}  \, .
      \label{domainQ}
\end{equation}
The space ${\cal D}(Q)$ is the domain of the position operator. Domain 
(\ref{domainQ}) is not the whole of $L^2$, since 
the function $g(x)=1/(x+\rmi)$ belongs to $L^2$ but not to
${\cal D}(Q)$; as well, $Q$ is an unbounded operator, because
$\| Qg \| = \infty$; as well, $Q{\cal D}(Q)$ is not included in 
${\cal D}(Q)$, since $h(x)=1/(x^2+1)$ belongs to ${\cal D}(Q)$ but $Qh$ does 
not belong to ${\cal D}(Q)$. The denseness and the non-invariance of the 
domains of unbounded operators create much trouble in the Hilbert space 
framework, because one has always to be careful whether formal operations
are valid. For example, $Q^2=QQ$ is not
defined on the whole of $L^2$, not even on the whole
of ${\cal D}(Q)$, but only on those square integrable functions such that
$x^2f \in L^2$. Also, the expectation value of the measurement of $Q$ in
the state $\varphi$,
\begin{equation}
       (\varphi ,Q\varphi ) \, ,  
        \label{exintrodispQ}
\end{equation}
is not finite for every $\varphi \in L^2$, but only when 
$\varphi \in {\cal D}(Q)$. Similarly, the uncertainty of the measurement
of $Q$ in $\varphi$,
\begin{equation}
      \Delta _{\varphi}Q=
      \sqrt{ (\varphi ,Q^2\varphi )-(\varphi ,Q\varphi )^2} \, , 
      \label{introdispQ}
\end{equation}
is not defined on the whole of $L^2$. 

On the other hand, if we denote the momentum observable by 
\begin{equation}
      Pf(x)=-\rmi \hbar \frac{\rmd}{\rmd x}f(x) \, , 
        \label{fdopp}
\end{equation}
then the product of $P$ and $Q$, $PQ$, is not defined everywhere in the 
Hilbert space, but only on those square integrable functions for which the 
quantity
\begin{equation}
      PQf(x) = -\rmi \hbar \frac{\rmd}{\rmd x}xf(x)=
           -\rmi \hbar \left( f(x)+ xf'(x) \right)
      \label{pqf} 
\end{equation}
makes sense and is square integrable. Obviously, $PQf$ makes
sense only when $f$ is differentiable, and $PQf$ remains in
$L^2$ only when $f$, $f'$ and $xf'$ are also in
$L^2$; thus, $PQ$ is not defined everywhere in
$L^2$ but only on those square integrable functions that
satisfy the aforementioned conditions. Similar domain concerns arise
in calculating the commutator of $P$ with $Q$.

As in the case of the position operator, the domain ${\cal D}(A)$ of an 
unbounded operator $A$ does not coincide with the whole of 
$\cal H$~\cite{RS84}, but is just a dense subspace of 
$\cal H$~\cite{DENSE}; also, in general ${\cal D}(A)$ does not remain 
invariant under the action of $A$, that is, $A{\cal D}(A)$ is not included 
in ${\cal D}(A)$. Such non-invariance makes expectation values,
\begin{equation}
       (\varphi ,A\varphi )  \, , 
        \label{exintrodispP}
\end{equation}
uncertainties,
\begin{equation}
      \Delta _{\varphi}A=
      \sqrt{ (\varphi ,A^2\varphi )-(\varphi ,A\varphi )^2} \, , 
      \label{introdispA}
\end{equation}
and algebraic operations such as commutation relations not well defined on 
the whole of the Hilbert space $\cal H$~\cite{INFENER}. Thus, when the 
position, momentum and energy operators $Q$, $P$, $H$ are unbounded, it is 
natural to seek a subspace $\mathbf \Phi$ of $\cal H$ on which all of these 
physical quantities can be calculated and yield meaningful, finite 
values. Because the reason why these quantities may not be well defined is 
that the domains of $Q$, $P$ and $H$
are not invariant under the action of these operators, the subspace 
$\mathbf \Phi$ must be such that
it remains invariant under the actions of $Q$, $P$ and $H$. This is why we
take as $\mathbf \Phi$ the intersection of the domains of all the powers
of $Q$, $P$ and $H$~\cite{ROBERTS}:
\begin{equation}
      {\mathbf \Phi} =\bigcap _{\Sb n,m=0 \\ A,B=Q,P,H \endSb}^{\infty} 
            {\cal D}(A^nB^m) \, .
      \label{maximalinvas}
\end{equation}
This space is known as the maximal invariant subspace of the algebra generated
by $Q$, $P$ and $H$, because it is the largest subdomain of the Hilbert space 
that remains invariant under the action of any power of $Q$, $P$ or $H$,
\begin{equation}
      A {\mathbf \Phi} \subset  {\mathbf \Phi} \, , \qquad A=Q,P,H \, .
\end{equation}
On $\mathbf \Phi$, all physical quantities such as 
expectation values and uncertainties can be associated well-defined, finite 
values, and algebraic operations such as the commutation relation 
(\ref{comuts}) are well defined. In addition, the elements of
$\mathbf \Phi$ are represented by smooth, continuous functions that
have a definitive value at each point, in contrast
to the elements of $\cal H$, which are represented by classes of functions 
which can vary arbitrarily on sets of zero Lebesgue measure.

Not only there are compelling reasons to shrink the Hilbert space $\cal H$ to
$\mathbf \Phi$, but, as we are going to explain now, there are also reasons
to enlarge $\cal H$ to the spaces ${\mathbf \Phi}^{\times}$ and 
${\mathbf \Phi}^{\prime}$ of Eqs.~(\ref{RHStIntro}) and 
(\ref{RHSpIntro}). When the spectrum of $A$ has a continuous 
part, prescriptions~(\ref{braeqeneintro}) and (\ref{cketequeintro}) associate a
bra $\langle a|$ and a ket $|a\rangle$ to each element $a$ of the 
continuous spectrum of $A$. Obviously, the bras $\langle a|$ and kets 
$|a\rangle$ are not in the Hilbert space~\cite{SNHS}, and therefore we 
need two linear spaces larger than the Hilbert space to accommodate 
them. It turns
out that the bras and kets acquire mathematical meaning as distributions. More 
specifically, the bras $\langle a|$ are {\it linear} functionals over
the space $\mathbf \Phi$, and the kets $|a\rangle$ are 
{\it antilinear} functionals over the space 
$\mathbf \Phi$. That is, $\langle a| \in {\mathbf \Phi}^{\prime}$ and 
$|a \rangle  \in {\mathbf \Phi}^{\times}$.

In this way, the Gelfand triplets of Eqs.~(\ref{RHStIntro}) and 
(\ref{RHSpIntro}) arise in a natural way. The Hilbert space $\mathcal H$ 
arises from the requirement that the wave functions be square 
normalizable. Aside from providing mathematical concepts such as 
self-adjointness or unitarity, the Hilbert space plays a very important 
physical role, namely $\cal H$ selects the scalar product that is used to 
calculate probability amplitudes. The subspace $\mathbf \Phi$ contains those 
square integrable functions that should be considered as physical, because 
any expectation value, any uncertainty and any algebraic operation can be 
calculated for its elements, whereas this is not possible for the rest of the 
elements of the Hilbert space. The dual space ${\mathbf \Phi}^{\prime}$ and 
the antidual space ${\mathbf \Phi}^{\times}$ contain respectively the bras 
and the kets associated with the continuous spectrum of the observables. These
bras and kets can be used to expand any $\varphi \in {\mathbf \Phi}$ as in 
Eq.~(\ref{introDirbaexp}). Thus, the rigged Hilbert space, rather than 
the Hilbert space alone, can accommodate prescriptions 
(\ref{dketqueintro})-(\ref{comuts}) of Dirac's formalism.

It should be clear that the rigged Hilbert space is just a combination of 
the Hilbert space with distribution theory. This combination enables us to 
deal with singular objects such as bras, kets, or Dirac's delta function, 
something that is impossible if we only use the Hilbert space.

Even though it is apparent that the rigged Hilbert space should be an 
essential part of the mathematical methods for Quantum Mechanics, one may
still wonder if the rigged Hilbert space is a helpful tool in teaching 
Quantum Mechanics, or rather is a technical nuance. Because basic quantum 
mechanical operators such as $P$ and $Q$ are in general unbounded operators 
with continuous spectrum~\cite{RS274}, and because this kind of operators 
necessitates the rigged Hilbert space, it seems pertinent to introduce the 
rigged Hilbert space in graduate courses on Quantum Mechanics.

From a pedagogical standpoint, however, this section's introduction to the 
rigged Hilbert space is not sufficient. In the classroom, new concepts are 
better introduced by way of a simple, exactly solvable example. This is why
we shall construct the RHS of the 1D rectangular barrier system. We note that
this system does not have bound states, and therefore in what follows 
we shall not deal with discrete spectrum.

\subsection{Representations}

In working out specific examples, the prescriptions of Dirac's 
formalism have to be written in a particular representation. Thus, before
constructing the RHS of the 1D rectangular barrier, it is convenient to
recall some of the basics of representations.

In Quantum Mechanics, the most common of all representations is the position
representation, sometimes called the $x$-representation. In the
$x$-representation, the position operator $Q$ acts as multiplication by 
$x$. Since the spectrum of $Q$ is $(-\infty ,\infty )$, the $x$-representation
of the Hilbert space $\cal H$ is given by the space 
$L^2$. In this paper, we shall mainly work in the
position representation.

In general, given an observable $B$, the $b$-representation is that in which 
the operator $B$ acts as multiplication by $b$, where the $b$'s denote
the eigenvalues of $B$. If we denote the spectrum of $B$ by ${\rm Sp}(B)$, then
the $b$-representation of the Hilbert space $\cal H$ is given by the 
space $L^2( {\rm Sp}(B),\rmd b)$, which is the space of square integrable 
functions $f(b)$ with $b$ running over ${\rm Sp}(B)$. In the 
$b$-representation, the restrictions to purely continuous spectrum of 
prescriptions~(\ref{dketqueintro})-(\ref{introresiden}) become
\numparts
    \begin{equation}
           \langle b|A|a \rangle =a \langle b|a \rangle \, ,
         \label{bqueintro}    
   \end{equation}
   \begin{equation}
         \langle a|A|b\rangle =a \langle a|b\rangle \, ,
          \label{bbraeqeneintro}
    \end{equation}
    \begin{equation}
      \langle b| \varphi \rangle = 
       \int \rmd a \, \langle b|a \rangle \langle a |\varphi \rangle \, ,
        \label{bintroDirbaexp}
      \end{equation}
   \begin{equation}
       \delta (b-b')=\langle b|b'\rangle = 
       \int \rmd a \, \langle b|a \rangle \langle a |b'\rangle \, .
        \label{bintroresiden}
      \end{equation}
\endnumparts 
The ``scalar product'' $\langle b|a\rangle$ is obtained from 
Eq.~(\ref{bqueintro}) as the solution of a differential eigenequation in 
the $b$-representation. The $\langle b|a\rangle$ can also be seen as 
transition elements from the 
$a$- to the $b$-representation. Mathematically, the $\langle b|a\rangle$  
are to be treated as distributions, and therefore they often appear
as kernels of integrals. In this paper, we shall encounter a few of these 
``scalar products'' such as $\langle x| p\rangle$, $\langle x| x'\rangle$
and $\langle x| E^{\pm} \rangle _{\rm l,r}$.

\section{Example: The one-dimensional rectangular barrier potential}
\label{sec:e1dsbp}

The example we consider in this paper is supposed to represent a spinless 
particle moving in one dimension and impinging on a rectangular barrier. The 
observables relevant to this system are the position $Q$, the momentum $P$, and
the Hamiltonian $H$. In the position representation, $Q$ and $P$ are 
respectively realized by the differential operators (\ref{fdopx}) and
(\ref{fdopp}), whereas $H$ is realized by
\begin{equation} 
      Hf(x)= \left( -\frac{\hbar ^2}{2m}\frac{\rmd ^2}{\rmd x^2}+V(x) \right) 
                  f(x) 
          \, , 
         \label{fdoph}
\end{equation}
where
\begin{equation}
           V(x)=\left\{ \begin{array}{ll}
                                0   &-\infty <x<a  \\
                                V_0 &a<x<b  \\
                                0   &b<x<\infty 
                  \end{array} 
                 \right. 
	\label{sbpotential}
\end{equation}
is the 1D rectangular barrier potential. Formally, these observables satisfy 
the following commutation relations:
\numparts
\begin{equation}
      \left[ Q,P \right] =\rmi \hbar I \, , \label{cr1} 
\end{equation}
\begin{equation}
      \left[ H,Q \right] =- \frac{\rmi \hbar}{m} P \, ,   
\end{equation}
\begin{equation}
      \left[ H,P \right] = \rmi \hbar \frac{\partial V}{\partial x} \, .  
        \label{cr3}
\end{equation}
\endnumparts

Since our particle can move in the full real line, the Hilbert space
on which the differential operators~(\ref{fdopx}), (\ref{fdopp}) and 
(\ref{fdoph}) should act is 
$L^2$ of Eq.~(\ref{l2space}). The corresponding scalar 
product is
\begin{equation}
      (f,g)=\int_{-\infty}^{\infty}\rmd x \, \overline{f(x)}g(x) \, , \qquad
      f,g \in L^2 \, ,
          \label{scapro}
\end{equation}
where $\overline{f(x)}$ denotes the complex conjugate of $f(x)$.

The differential operators~(\ref{fdopx}), (\ref{fdopp}) and (\ref{fdoph})
induce three linear
operators on the Hilbert space $L^2$. These operators are 
unbounded~\cite{FOCO}, and therefore they cannot be defined on the whole of 
$L^2$, but only on the following subdomains of
$L^2$~\cite{FOCO}:
\numparts
\begin{equation}
     {\cal D}(Q)=\left\{ f\in L^2 \, | \ 
                   xf \in  L^2 \right\} \, ,  
\end{equation}
\begin{equation}
     {\cal D}(P)=\left\{ f\in L^2 \, | \ 
                f \in  AC, \
                Pf \in  L^2 \right\} \, , 
\end{equation}
\begin{equation}
     {\cal D}(H)=\left\{ f\in L^2 \, | \ 
                f \in  AC^2, \
                Hf \in  L^2 \right\} \, , 
         \label{domainH} 
\end{equation}
\endnumparts
where, essentially, $AC$ is the space of functions whose 
derivative exists, and $AC^2$ is the space of functions whose
second derivative exists (see Ref.~\cite{FOCO} for more details). On 
these domains, the operators $Q$, $P$ and $H$ are 
self-adjoint~\cite{FOCO}. 

In our example, the eigenvalues (i.e., the spectrum) and the eigenfunctions
of the observables are provided by the Sturm-Liouville theory. Mathematically,
the eigenvalues and eigenfunctions of operators extend the notions of 
eigenvalues and eigenvectors of a matrix to the infinite-dimensional case. The
Sturm-Liouville theory tells us that these operators have the 
following spectra~\cite{FOCO}:
\numparts
\begin{equation}
      {\rm Sp}(Q)=(-\infty, \infty) \, ,
\end{equation}
\begin{equation}
      {\rm Sp}(P)=(-\infty, \infty) \, , 
\end{equation}
\begin{equation}
      {\rm Sp}(H)=[0, \infty) \, . 
\end{equation}
\endnumparts
These spectra coincide with those we would expect on physical grounds. We
expect the possible measurements of $Q$ to be the full real line, because
the particle can in principle reach any point of the real line. We also 
expect the possible measurements of $P$ to be the full real line, since the 
momentum of the particle is not restricted in magnitude or direction. The 
possible measurements of $H$ have the same range as that of the kinetic 
energy, because the potential does not have any wells of negative energy,
and therefore we expect the spectrum of $H$ to be the positive real line.

To obtain the eigenfunction corresponding to each eigenvalue, we have to 
solve the eigenvalue equation~(\ref{cketequeintro}) for each observable. Since 
we are working in the position representation, we have to write 
Eq.~(\ref{cketequeintro}) in the position representation for each observable:
\numparts
\begin{equation}
    \langle x|Q|x'\rangle = x' \langle x|x'\rangle \, ,  
       \label{xqee}
\end{equation}
\begin{equation}
    \langle x|P|p\rangle = p \langle x|p\rangle \, , 
    \label{xpee}
\end{equation}
\begin{equation}
    \langle x|H|E\rangle = E \langle x|E\rangle \, .
    \label{xhee} 
\end{equation}
\endnumparts
By recalling Eqs.~(\ref{fdopx}), (\ref{fdopp}) and (\ref{fdoph}), we can 
write Eqs.~(\ref{xqee})-(\ref{xhee}) as
\numparts
\begin{equation}
      x \langle x|x'\rangle = x' \langle x|x'\rangle \, , 
         \label{eeq}
\end{equation}
\begin{equation} 
      -\rmi \hbar \frac{\rmd}{\rmd x} \langle x|p\rangle = 
              p \langle x|p\rangle  \, ,
       \label{eep}
\end{equation}
\begin{equation} 
     \left( -\frac{\hbar ^2}{2m}\frac{\rmd ^2}{\rmd x^2}    
        +V(x) \right) \langle x|E\rangle = E \langle x|E\rangle \, . 
         \label{eeh}
\end{equation}
\endnumparts
For each position $x'$, Eq.~(\ref{eeq}) yields the corresponding eigenfunction
of $Q$ as a delta function,
\begin{equation}
      \langle x| x'\rangle = \delta (x-x') \, .
       \label{deltax}
\end{equation}
For each momentum $p$, Eq.~(\ref{eep}) yields the corresponding eigenfunction
of $P$ as a plane wave,
\begin{equation}
      \langle x| p\rangle = \frac{\rme ^{\rmi px/\hbar}}{\sqrt{2\pi \hbar}} 
                   \, .
       \label{expp}
\end{equation}
For each energy $E$, Eq.~(\ref{eeh}) yields the following two linearly 
independent eigenfunctions~\cite{FOCO}:
\numparts
\begin{equation}
           \langle x|E^+\rangle _{\rm r}   = 
          \left( \frac{m}{2\pi k \hbar ^2} \right)^{1/2} \times   
                \left\{ \begin{array}{lc}
             T (k)\rme ^{-\rmi kx}  &-\infty <x<a  \\
             A_{\rm r}(k)\rme ^{\rmi \kappa x}+
                       B_{\rm r}(k) \rme^{-\rmi \kappa x} &a<x<b \\
        R_{\rm r}(k)\rme ^{\rmi kx} + \rme ^{-\rmi kx}   &b<x<\infty \, ,
                  \end{array} 
                 \right. 
     \label{chir+}
\end{equation}
\begin{equation}
         \langle x|E^+\rangle _{\rm l}=          
        \left( \frac{m}{2\pi k \hbar ^2} \right)^{1/2}
               \times  \left\{ \begin{array}{lc}
             \rme ^{\rmi kx}+R_{\rm l}(k)\rme ^{-\rmi kx}  &-\infty <x<a  \\
     A_{\rm l}(k)\rme ^{\rmi \kappa x}+B_{\rm l}(k)\rme ^{-\rmi \kappa x}
                                   &a<x<b \\
               T (k)\rme ^{\rmi kx}         &b<x<\infty \, ,
                  \end{array} 
                 \right.
          \label{chil+} 
\end{equation}
\endnumparts
where
\begin{equation}
      k=\sqrt{\frac{2m}{\hbar ^2}E} \, , \quad
      \kappa =\sqrt{\frac{2m}{\hbar ^2}(E-V_0)} \, ,
\end{equation}
and where the coefficients that appear in Eqs.~(\ref{chir+})-(\ref{chil+}) can 
be easily found by the standard matching conditions at the discontinuities of 
the potential~\cite{FOCO}. Thus, in contrast to the spectra of $Q$ and $P$, 
the spectrum of $H$ is doubly degenerate.

\begin{figure}
\begin{center}
\epsfbox{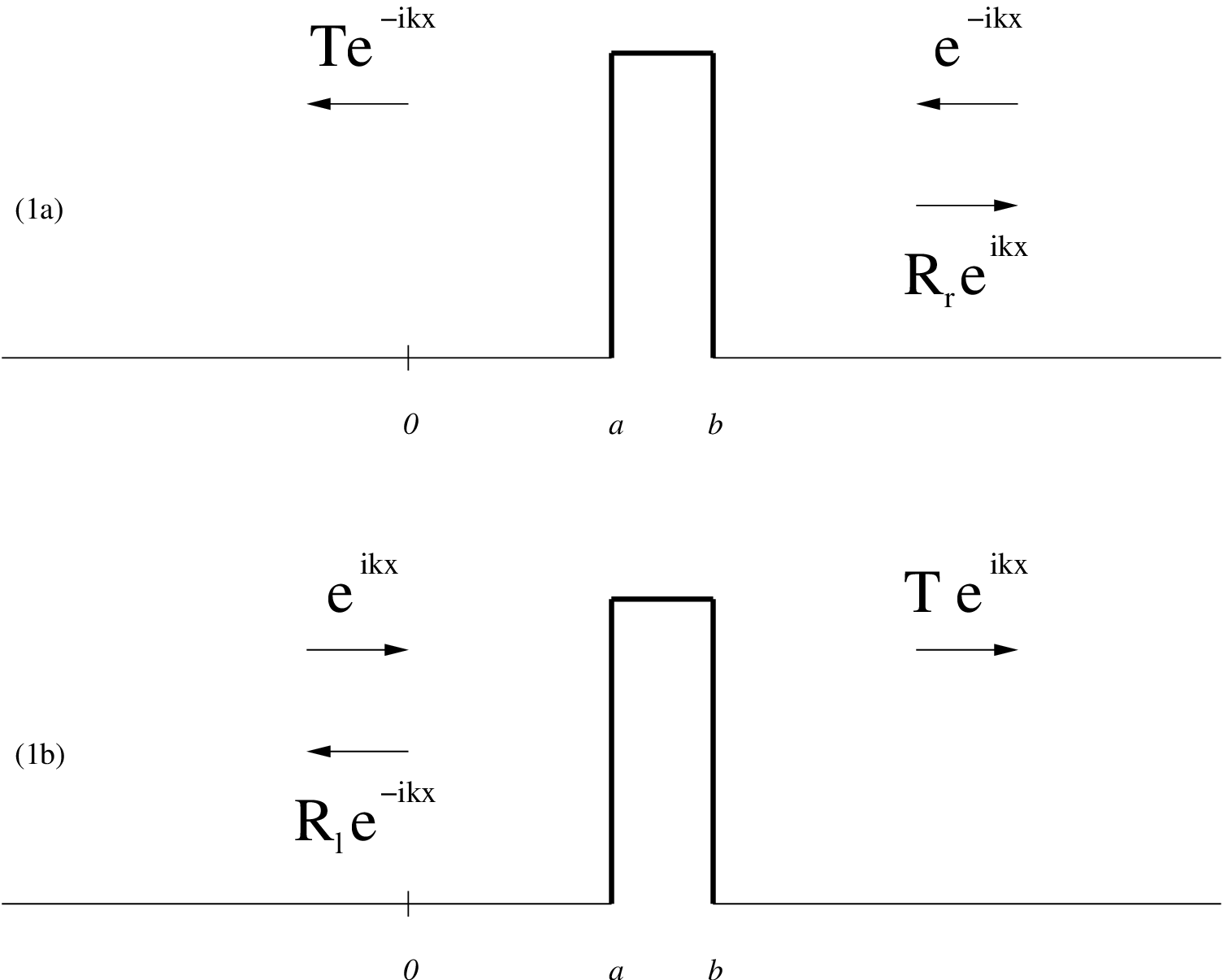}
\end{center}
\caption{Schematic representation of the eigenfunctions 
$\langle x|E^+\rangle _{\rm r}$, Fig.~1a, and 
$\langle x|E^+\rangle _{\rm l}$, Fig.~1b.}
\label{fig:plus}
\end{figure}

Physically,
the eigenfunction $\langle x|E^+\rangle _{\rm r}$ represents a particle of
energy $E$ that impinges on the barrier from the right (hence the subscript r)
and gets reflected to the right with probability amplitude $R_{\rm r}(k)$ and 
transmitted to the left with probability amplitude $T (k)$, see
Fig.~\ref{fig:plus}a. The eigenfunction
$\langle x|E^+\rangle _{\rm l}$ represents a particle of energy $E$ that 
impinges on the barrier from the left (hence the subscript l) and gets 
reflected to the left with probability amplitude $R_{\rm l}(k)$ and 
transmitted to the right with probability amplitude $T(k)$, see 
Fig.~\ref{fig:plus}b. 

\begin{figure}
\begin{center}
\epsfbox{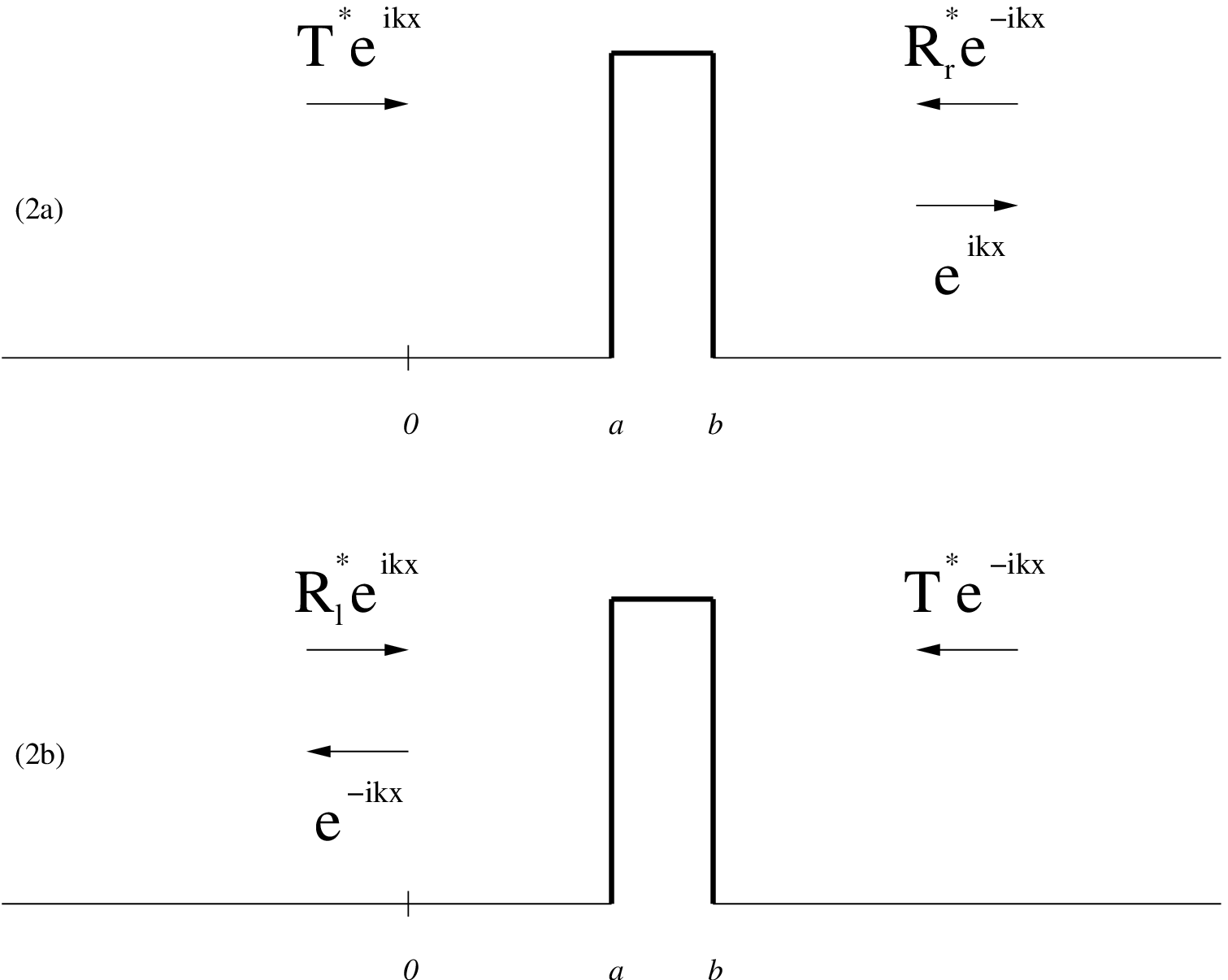}
\end{center}
\caption{Schematic representation of the eigenfunctions 
$\langle x|E^-\rangle _{\rm r}$, Fig.~2a, and 
$\langle x|E^-\rangle _{\rm l}$, Fig.~2b.}
\label{fig:minus}
\end{figure}

Note that, instead of~(\ref{chir+})-(\ref{chil+}), we could choose another 
pair of linearly independent solutions of Eq.~(\ref{eeh}) as 
follows~\cite{FOCO}:
\numparts
\begin{equation}
           \langle x|E^-\rangle _{\rm r} = 
         \left( \frac{m}{2\pi k \hbar ^2} \right)^{1/2}
              \times \left\{ \begin{array}{lc}
             T^*(k)\rme ^{\rmi kx}  &-\infty <x<a  \\
         A_{\rm r}^*(k)\rme ^{-\rmi \kappa x}+
                    B_{\rm r}^*(k) \rme ^{\rmi \kappa x}&a<x<b \\
        R_{\rm r}^*(k)\rme ^{-\rmi kx} + \rme ^{\rmi kx}   &b<x<\infty \, ,
                  \end{array} 
                 \right. 
             \label{chir-}
\end{equation}
\begin{equation}
          \langle x|E^-\rangle _{\rm l}     = 
         \left( \frac{m}{2\pi k \hbar ^2} \right)^{1/2}
              \times \left\{ \begin{array}{lc}
         \rme ^{-\rmi kx}+R_{\rm l}^*(k)\rme ^{\rmi kx}  &-\infty <x<a  \\
         A_{\rm l}^*(k)\rme ^{-\rmi \kappa x}+
                 B_{\rm l}^*(k)\rme ^{\rmi \kappa x}&a<x<b \\
               T^*(k)\rme ^{-\rmi kx}         &b<x<\infty \, ,
                  \end{array} 
                 \right.
          \label{chil-} 
\end{equation}
\endnumparts
where the coefficients of these eigenfunctions can also be calculating
by means of the standard matching conditions at $x=a,b$~\cite{FOCO}. The 
eigenfunction $\langle x|E^-\rangle _{\rm r}$ 
represents two plane waves---one impinging on the barrier from the left with 
probability amplitude $T^*(k)$ and another impinging on the barrier from the
right with probability amplitude $R_{\rm r}^*(k)$---that combine in such a way 
as to produce an outgoing plane wave to the right, see 
Fig.~\ref{fig:minus}a. The eigenfunction
$\langle x|E^-\rangle _{\rm l}$ represents two other planes waves---one 
impinging on the barrier from left with probability amplitude $R_{\rm l}^*(k)$ 
and another impinging on the 
barrier from the right with probability amplitude $T^*(k)$---that combine in 
such a way as to produce an outgoing wave to the left, see 
Fig.~\ref{fig:minus}b. The eigensolutions
$\langle x|E^-\rangle _{\rm r,l}$ correspond to the {\it final} condition 
of an outgoing plane wave propagating away from the barrier respectively to 
the right and to the left, as opposed to $\langle x|E^+\rangle _{\rm r,l}$, 
which correspond to the {\it initial} condition of a plane wave that 
propagates towards the barrier respectively from the right and from the left.

The eigenfunctions (\ref{deltax}), (\ref{expp}), (\ref{chir+})-(\ref{chil+}) 
and (\ref{chir-})-(\ref{chil-})
are not square integrable, that is, they do not belong to 
$L^2$. Mathematically speaking, this is the reason why 
they are
to be dealt with as distributions (note that all of them except for the delta
function are also proper functions). Physically speaking, they
are to be interpreted in analogy to electromagnetic plane waves, as we shall
see in Section~\ref{sec:phymean}.

\section{Construction of the rigged Hilbert space}
\label{sec:crhs}

In the previous section, we saw that the observables of our system are 
implemented by unbounded operators with continuous spectrum. We also saw
that the eigenfunctions of the observables do not belong to 
$L^2$. Thus, as we explained in Sec.~\ref{sec:why}, we need
to construct the rigged Hilbert spaces of Eqs.~(\ref{RHStIntro}) 
and (\ref{RHSpIntro}) [see Eqs.~(\ref{RHSCONTpr}) and (\ref{RHSCONTprb}) 
below]. We start by constructing $\mathbf \Phi$.

\subsection{Construction of $\mathbf \Phi \equiv \Sw$}

The subspace $\mathbf \Phi$ is given by Eq.~(\ref{maximalinvas}). In view of 
expressions (\ref{fdopx}), (\ref{fdopp}) and (\ref{fdoph}), the elements of 
$\mathbf \Phi$ must fulfill the following conditions:
\begin{itemize}
  \item[$\bullet$] they are infinitely differentiable, so the differentiation
operation can be applied as many times as wished,
  \item[$\bullet$] they vanish at $x=a$ and $x=b$, so differentiation is 
meaningful at the discontinuities of the potential~\cite{ZERODERIV},
  \item[$\bullet$] the action of all powers of $Q$, $P$ and $H$ remains square
integrable.
\end{itemize}
Hence,
\begin{eqnarray}
   \hskip-1cm {\mathbf \Phi} =\{ \varphi \in L^2 \, | \
    \varphi \in C^{\infty}(\mathbb R), \ \varphi ^{(n)}(a)=\varphi ^{(n)}(b)=0 
    \, , \ n=0,1,\ldots \, ,  \nonumber \\
    \hskip3.6cm
     P^nQ^mH^l\varphi (x) \in L^2
    \, , \   n,m,l=0,1, \ldots  \} \, ,
     \label{ddomain}
\end{eqnarray}
where $C^{\infty}(\mathbb R)$ is the collection of infinite differentiable
functions, and $\varphi ^{(n)}$ denotes the $n$th derivative of 
$\varphi$. From the last condition in Eq.~(\ref{ddomain}), we deduce that the 
elements of $\mathbf \Phi$ satisfy the following estimates:
\begin{equation}
      \| \varphi \| _{n,m,l} \equiv  
 \sqrt{\int_{-\infty}^{\infty}\rmd x 
          \, \left| P^nQ^mH^l\varphi (x)\right| ^2 \,}
   < \infty  \, , \quad n,m,l=0,1,\ldots \, .
      \label{nmnorms}
\end{equation}
These estimates mean that the action of any combination of any power of the
observables remains square integrable. For this to happen, the functions
$\varphi (x)$ must be infinitely differentiable and must fall off at infinity
faster than any polynomial. The estimates~(\ref{nmnorms}) induce a topology
on $\mathbf \Phi$, that is, they induce a meaning of convergence of 
sequences, in the following way. A sequence $\{ \varphi _{\alpha} \}$ 
$\mathbf \Phi$-converges to $\varphi$ when $\{ \varphi _{\alpha} \}$ converges
to $\varphi$ with respect to all the estimates~(\ref{nmnorms}),
\begin{equation}
      \varphi _{\alpha}\, 
       \mapupdown{\tau_{\mathbf \Phi}}{\alpha \to \infty}
      \, \varphi \quad {\rm if} \quad  
      \| \varphi _{\alpha }-\varphi \| _{n,m,l} 
      \, \mapupdown{}{\alpha \to \infty}\, 0 \, , \quad n,m,l=0,1, \ldots \, .
\end{equation} 
Intuitively, a sequence $\varphi _{\alpha}$ converges to $\varphi$ if
whenever we follow the terms of the sequence, we get closer and closer to
the limit point $\varphi$ with respect to a certain sense of closeness. In our
system, the notion of closeness is determined by the estimates 
$\|  \ \|_{n,m,l}$, which originate from the physical
requirements that led us to construct $\mathbf \Phi$.

From Eqs.~(\ref{ddomain}) and (\ref{nmnorms}), we can see that $\mathbf \Phi$ 
is very similar to the Schwartz space ${\cal S}(\mathbb R)$, the
major differences being that the derivatives of the elements of $\mathbf \Phi$ 
vanish at $x=a,b$ and that $\mathbf \Phi$ is not only invariant under
$P$ and $Q$ but also under $H$. This is why we shall write
\begin{equation}
       \mathbf \Phi \equiv \Sw \, .
\end{equation}

It is always a good, though lengthy exercise to check that $\Sw$ is indeed 
invariant under the action of the observables,
\begin{equation}
      A \, \Sw \subset \Sw \, , \qquad A=P,Q,H \, .
\end{equation}
This invariance guarantees that the expectation values
\begin{equation}
      (\varphi , A^n\varphi ) \, , \quad \varphi \in \Sw \, ,
      \ A=P, Q, H , \ n=0,1,\ldots 
\end{equation}
are finite, and that the commutation relations~(\ref{cr1})-(\ref{cr3}) are 
well 
defined~\cite{CRVANISHES}. It can also be checked that $P$, $Q$ and $H$, which 
are not continuous
with respect the topology of the Hilbert space $L^2$, are 
now continuous with respect to the topology $\tau _{\mathbf \Phi}$ of 
$\Sw$~\cite{FOCO,DIS}.

\subsection{Construction of $\mathbf \Phi ^{\times}\equiv \Swt$. The Dirac 
kets}

The space $\mathbf \Phi ^{\times}$ is simply the collection of 
$\tau _{\mathbf \Phi}$-continuous {\it antilinear} functionals over 
$\mathbf \Phi$~\cite{FUNCTIONAL}. By combining the spaces $\mathbf \Phi$,
$\cal H$ and $\mathbf \Phi ^{\times}$, we obtain the RHS of our system,
\begin{equation}
       \mathbf \Phi \subset {\cal H}\subset \mathbf \Phi ^{\times}
        \, ,
       \label{RHSCONT}
\end{equation}
which we denote in the position representation by
\begin{equation}
      \rhsSwt     \, .
       \label{RHSCONTpr}
\end{equation}
The space $\Swt$ is meant to accommodate the eigenkets $|p \rangle$, 
$|x\rangle$ and $|E^{\pm}\rangle _{\rm l,r}$ of $P$, $Q$ and $H$. In the 
remainder of this subsection, we construct these eigenkets explicitly and see 
that they indeed belong to $\Swt$. We shall also see
that $|p \rangle$, $|x\rangle$ and $|E^{\pm}\rangle _{\rm l,r}$ are indeed
eigenvectors of the observables.

The definition of a ket is borrowed from the theory of
distributions as follows~\cite{GELFAND}. Given a function $f(x)$ and a space 
of test functions $\mathbf \Phi$, the antilinear functional $F$ that 
corresponds to the function $f(x)$ is an integral operator whose kernel is 
precisely $f(x)$:
\numparts
\begin{equation}
      F(\varphi)=\int \rmd x \, \overline{\varphi (x)} f(x) \, ,
         \label{afFtff1}
\end{equation}
which in Dirac's notation becomes
\begin{equation}
      \langle \varphi|F\rangle =\int \rmd x \, \langle \varphi|x\rangle 
                                     \langle x|f\rangle \, . 
       \label{afFtff2}
\end{equation}
\endnumparts
It is important to keep in mind that, though related, the function $f(x)$ and
the functional $F$ are two different things, the relation between them being 
that $f(x)$ is the kernel of $F$ when we write $F$ as an integral operator. In
the physics literature, the term {\it distribution} is usually reserved 
for $f(x)$.

Definition~(\ref{afFtff1}) provides the link between the quantum mechanical
formalism and the theory of distributions. In practical applications, what
one obtains from the quantum mechanical formalism is the distribution
$f(x)$ (in this paper, the plane waves
$\frac{1}{\sqrt{2\pi \hbar}} \rme ^{\rmi px/\hbar}$, the delta function
$\delta (x-x')$ and the eigenfunctions 
$\langle x|E^{\pm}\rangle _{\rm l,r}$). Once $f(x)$ is given, one
can use definition~(\ref{afFtff1}) to generate the functional 
$|F\rangle$. Then,
the theory of distributions can be used to obtain the properties of the 
functional $|F\rangle$, which in turn yield the properties of the distribution
$f(x)$. 

By using prescription~(\ref{afFtff1}), we can define for 
each eigenvalue $p$ the eigenket $|p\rangle$ associated with the 
eigenfunction (\ref{expp}):
\numparts
\begin{equation}
       \langle \varphi |p\rangle  \equiv 
       \int_{-\infty}^{\infty}\rmd x \, \overline{\varphi (x)}
          \frac{1}{\sqrt{2\pi \hbar}} \rme ^{\rmi px/\hbar}  \, ,
     \label{definitionketp}
\end{equation}
which, using Dirac's notation for the integrand, becomes 
\begin{equation}
       \langle \varphi |p\rangle  \equiv 
       \int_{-\infty}^{\infty}\rmd x \, \langle \varphi |x\rangle 
       \langle x|p\rangle \, .
     \label{definitionketpDirac}
\end{equation}
\endnumparts
Similarly, for each $x$, we can define the ket $|x\rangle$ associated with
the eigenfunction (\ref{deltax}) of the position operator as
\numparts
\begin{equation}
     \langle \varphi |x\rangle  \equiv 
       \int_{-\infty}^{\infty}\rmd x' \, \overline{\varphi (x')}
          \delta (x-x') \, ,
     \label{definitionketx}
\end{equation}
which, using Dirac's notation for the integrand, becomes
\begin{equation}
     \langle \varphi |x\rangle  \equiv 
       \int_{-\infty}^{\infty}\rmd x' \, \langle \varphi | x'\rangle
          \langle x'|x\rangle \, .
     \label{definitionketxDirac}
\end{equation}
\endnumparts
The definition of the kets $|E^{\pm}\rangle _{\rm l,r}$ that correspond to the 
Hamiltonian's eigenfunctions~(\ref{chir+})-(\ref{chil+}) and 
(\ref{chir-})-(\ref{chil-}) 
follows the same prescription:
\numparts
\begin{equation}
     \langle \varphi |E^{\pm}\rangle _{\rm l,r}  \equiv 
       \int_{-\infty}^{\infty}\rmd x \, \overline{\varphi (x)}
        \langle x|E^{\pm}\rangle _{\rm l,r}  \, ,
    \label{definitionketE}
\end{equation}
that is, 
\begin{equation}
     \langle \varphi |E^{\pm}\rangle _{\rm l,r}  \equiv 
       \int_{-\infty}^{\infty}\rmd x \, \langle \varphi |x\rangle
          \langle x|E^{\pm}\rangle _{\rm l,r} \, .
    \label{definitionketEDirac}
\end{equation}
\endnumparts
(Note that this equation defines four different 
kets.) One can now show that the definition of the kets $|p\rangle$, 
$|x\rangle$ and $|E^{\pm}\rangle _{\rm l,r}$ makes sense, and that
these kets indeed belong to the space of distributions $\Swt$~\cite{FOCO}.

As in the general case of Eqs.~(\ref{afFtff1})-(\ref{afFtff2}),
it is important to keep in mind the difference between eigenfunctions and
kets. For instance, $\langle x|p\rangle$ is an eigenfunction
of a differential equation, Eq.~(\ref{eep}), whereas $|p\rangle$ is a 
functional, the relation between them being given by 
Eq.~(\ref{definitionketpDirac}). A similar relation holds between 
$\langle x'|x\rangle$ and $|x\rangle$, and between 
$\langle x|E^{\pm}\rangle _{\rm l,r}$ and $|E^{\pm}\rangle _{\rm l,r}$. It
is also important to keep in mind that ``scalar products'' like
$\langle x|p\rangle$, $\langle x'|x\rangle$ or 
$\langle x|E^{\pm}\rangle _{\rm l,r}$ do not represent an actual scalar 
product of two functionals; these ``scalar products'' are simply solutions
to differential equations.

We now turn to the question of whether the kets $|p\rangle$, $|x\rangle$ and 
$|E^{\pm}\rangle_{\rm l,r}$ are eigenvectors of the corresponding 
observable [see Eqs.~(\ref{kpeP})-(\ref{kEeH}) below]. Since the
observables act in principle only on their Hilbert space domains, and since
the kets lie outside the Hilbert space, we need to extend the definition
of the observables from $\mathbf \Phi$ into $\mathbf \Phi ^{\times}$, in order
to specify how the observables act on the kets. The theory of distributions 
provides us with a precise prescription of how an observable acts on 
$\mathbf \Phi ^{\times}$, and therefore of how it acts on the kets, as 
follows~\cite{GELFAND}. The action of a self-adjoint operator $A$ on a 
functional $|F\rangle \in {\mathbf \Phi}^{\times}$ is defined as
\begin{equation}
     \langle \varphi |A|F\rangle \equiv \langle A\varphi |F\rangle 
      \, , \quad \mbox{for all} \ \varphi \ \mbox{in} \ \mathbf \Phi  \, .
      \label{adualext}
\end{equation}
Note that this definition extends the Hilbert space definition of a 
self-adjoint operator,
\begin{equation}
     (f,Ag)=(Af,g)  \, ,
      \label{saosr}
\end{equation}
which is valid only when $f$ and $g$ belong to the domain of 
$A$. In turn, Eq.~(\ref{adualext}) can be used to define the notion
of eigenket of an observable: A functional $|a\rangle$ in 
$\mathbf \Phi ^{\times}$ is an eigenket of $A$ with eigenvalue $a$ if
\begin{equation}
     \langle \varphi |A|a\rangle = \langle A \varphi |a\rangle =
     a \langle \varphi |a\rangle \, , \quad 
    \mbox{for all} \ \varphi \ \mbox{in} \ \mathbf \Phi  \, .
       \label{eeRHS}
\end{equation}
When the ``left sandwiching'' of this equation with the elements of 
$\mathbf \Phi$ is understood and therefore omitted, we shall simply write
\begin{equation}
              A|a\rangle = a |a\rangle \, ,
          \label{eeRHSws}
\end{equation}
which is just Dirac's eigenket equation (\ref{cketequeintro}). Thus,
Dirac's eigenket equation acquires a precise meaning through
Eq.~(\ref{eeRHS}), in the sense that it has to be understood as 
``left sandwiched'' with the wave functions $\varphi$ of $\mathbf \Phi$. 

By using definition~(\ref{eeRHS}), one can show that $|p\rangle$, $|x\rangle$ 
and $|E^{\pm}\rangle_{\rm l,r}$ are indeed eigenvectors of $P$, $Q$ and $H$, 
respectively~\cite{FOCO}:
\begin{equation}
       P|p\rangle=p|p\rangle \, , \quad p\in \mathbb R \, ,
        \label{kpeP}
\end{equation}
\begin{equation}
       Q|x\rangle=x|x\rangle \, , \quad x \in \mathbb R \, ,
\end{equation}
\begin{equation}
       H|E^{\pm}\rangle_{\rm l,r} =E|E^{\pm}\rangle _{\rm l,r}
   \, , \quad E\in [0,\infty )  \, .
        \label{kEeH}
\end{equation}

\subsection{Construction of $\mathbf \Phi ^{\prime}\equiv \Swp$. The Dirac 
bras}

In complete analogy with the construction of the Dirac kets, we construct in 
this subsection the Dirac bras $\langle p|$, $\langle x|$ and 
$_{\rm l,r}\langle ^{\pm}E|$
of $P$, $Q$ and $H$. Mathematically, the Dirac bras are distributions that 
belong to the space $\mathbf \Phi ^{\prime}$, which is the space of 
{\it linear} functionals over $\mathbf \Phi$~\cite{FUNCTIONAL}. The 
corresponding RHS is
\begin{equation}
      {\mathbf \Phi} \subset {\cal H} \subset {\mathbf \Phi}^{\prime} \, ,
\end{equation}
which we denote in the position representation by
\begin{equation}
      \Sw \subset L^2 \subset \Swp \, .
       \label{RHSCONTprb}
\end{equation}

Likewise the definition of a ket, the definition of a bra is borrowed from
the theory of distributions~\cite{GELFAND}. Given a function $f(x)$ and
a space of test functions $\mathbf \Phi$, the linear functional $\tilde{F}$
generated by the function $f(x)$ is an integral operator whose kernel
is the complex conjugate of $f(x)$:
\numparts
\begin{equation}
      \tilde{F}(\varphi)=\int \rmd x \, \varphi (x) \overline{f(x)} \, ,
          \label{afGtff1}
\end{equation}
which in Dirac's notation becomes
\begin{equation}
      \langle F|\varphi \rangle =\int \rmd x \, \langle f|x\rangle
                                            \langle x|\varphi \rangle \, . 
\end{equation}
\endnumparts
Note that this definition is very similar to that of a linear functional,
Eq.~(\ref{afFtff1}), except that the complex conjugation affects $f(x)$ 
rather than $\varphi (x)$, which makes $\tilde{F}$ linear rather than 
antilinear. Likewise the 
antilinear case~(\ref{afFtff1}), it is important to keep in mind that, though 
related, the function $f(x)$ and the functional $\tilde{F}$ are two different 
objects, the relation between them being that $\overline{f(x)}$ is the kernel 
of $\tilde{F}$ when we write $\tilde{F}$ as an integral operator.

By using prescription~(\ref{afGtff1}), we can now define for each eigenvalue
$p$ the eigenbra $\langle p|$ associated with the eigenfunction~(\ref{expp}):
\numparts
\begin{equation}
    \langle p| \varphi \rangle  \equiv 
       \int_{-\infty}^{\infty}\rmd x \, \varphi (x)
          \frac{1}{\sqrt{2\pi \hbar}} \rme ^{-\rmi px/\hbar} \, ,
     \label{definitionbrap}
\end{equation}
which, using Dirac's notation for the integrand, becomes 
\begin{equation}
    \langle p| \varphi \rangle  \equiv 
       \int_{-\infty}^{\infty}\rmd x \, \langle p|x \rangle 
        \langle x| \varphi \rangle  \, .
     \label{definitionbrapDirac}
\end{equation}
\endnumparts
Comparison with Eq.~(\ref{definitionketp}) shows that the action of
$\langle p|$ is the complex conjugate of the action of $|p \rangle$,
\begin{equation}
      \langle p| \varphi \rangle = \overline{\langle \varphi |p \rangle} \, ,
\end{equation} 
and that
\begin{equation}
      \langle p| x \rangle = \overline{\langle x |p \rangle} =
      \frac{1}{\sqrt{2\pi \hbar}} \rme ^{-\rmi px/\hbar}  \, .
    \label{pxiscmplxp}
\end{equation} 
The bra $\langle x|$ is defined as
\numparts
\begin{equation}
     \langle x| \varphi \rangle  \equiv 
       \int_{-\infty}^{\infty}\rmd x' \, \varphi (x')
          \delta (x-x') \, ,
     \label{definitionbrax}
\end{equation}
which, using Dirac's notation for the integrand, becomes
\begin{equation}
     \langle x|\varphi \rangle  \equiv 
       \int_{-\infty}^{\infty}\rmd x' \, \langle x | x'\rangle
          \langle x'| \varphi \rangle \, .
     \label{definitionbraxDirac}
\end{equation}
\endnumparts
Comparison with Eq.~(\ref{definitionketx}) shows that the action of
$\langle x|$ is complex conjugated to the action of $|x \rangle$,
\begin{equation}
      \langle x| \varphi \rangle = \overline{\langle \varphi |x \rangle} \, ,
\end{equation} 
and that
\begin{equation}
      \langle x| x' \rangle = \langle x' |x \rangle = \delta (x-x')  \, .
\end{equation} 
Analogously, the eigenbras of the Hamiltonian are defined as
\numparts
\begin{equation}
     _{\rm l,r}\langle ^{\pm}E|\varphi\rangle \equiv
       \int_{-\infty}^{\infty}\rmd x \ \varphi (x) \
           _{\rm l,r} \langle ^{\pm}E|x\rangle \, ,
    \label{definitionbraE}
\end{equation}
that is, 
\begin{equation}
       _{\rm l,r}\langle ^{\pm}E|\varphi\rangle \equiv
       \int_{-\infty}^{\infty}\rmd x \ 
       _{\rm l,r}\langle ^{\pm}E|x\rangle \langle x|\varphi \rangle   \, ,
    \label{definitionbraEDirac}
\end{equation}
\endnumparts
where
\begin{equation}
     _{\rm l,r}\langle ^{\pm}E|x\rangle = 
   \overline{\langle x|E^{\pm}\rangle}_{\rm l,r} \, .
\end{equation}
(Note that in Eq.~(\ref{definitionbraE}) we have defined four different 
bras.) Comparison of Eq.~(\ref{definitionbraE}) with 
Eq.~(\ref{definitionketE}) shows that 
the actions of the bras $_{\rm l,r}\langle ^{\pm}E|$ are the complex 
conjugates of the actions of the kets $|E ^{\pm} \rangle _{\rm l,r}$:
\begin{equation}
      _{\rm l,r}\langle ^{\pm}E|\varphi\rangle =      
      \overline{\langle \varphi |E ^{\pm} \rangle}_{\rm l,r} \, .
      \label{braketccE}
\end{equation} 
Now, by using the RHS mathematics, one can show that the definitions of 
$\langle p|$, $\langle x|$ and $_{\rm l,r} \langle ^{\pm}E|$ make sense
and that $\langle p|$, $\langle x|$ and $_{\rm l,r} \langle ^{\pm}E|$
belong to $\Swp$~\cite{FOCO}.

Our next task is to see that the bras we just defined are left eigenvectors of 
the corresponding observable [see Eqs.~(\ref{bpeP})-(\ref{bEeH}) below]. For
this purpose, we need to specify how the observables act on the bras, 
that is, how they act on the dual space $\Swp$. We shall do so in analogy to 
the definition of their action on the kets, by means of the theory of 
distributions~\cite{GELFAND}. The action to the left of a self-adjoint 
operator $A$ on a linear functional $\langle F| \in {\mathbf \Phi}^{\prime}$ 
is defined as
\begin{equation}
     \langle F|A|\varphi \rangle \equiv \langle F| A\varphi \rangle 
      \, , \quad \mbox{for all} \ \varphi \ \mbox{in} \ \mathbf \Phi  \, .
      \label{dualext}
\end{equation}
Likewise definition~(\ref{adualext}), this definition generalizes 
Eq.~(\ref{saosr}). In turn, Eq.~(\ref{dualext}) can be used to define 
the notion of eigenbra of an observable: A functional $\langle a|$ in 
$\mathbf \Phi ^{\prime}$ is an eigenbra of $A$ with eigenvalue $a$ if
\begin{equation}
     \langle a |A|\varphi \rangle = \langle a|A \varphi \rangle =
     a \langle a|\varphi \rangle \, , \quad 
    \mbox{for all} \ \varphi \ \mbox{in} \ \mathbf \Phi  \, .
       \label{eeRHSb}
\end{equation}
When the ``right sandwiching'' of this equation with the elements of 
$\mathbf \Phi$ is understood and therefore omitted, we shall simply write
\begin{equation}
           \langle a |A = a \langle a|  \, ,
         \label{eeRHSwsb}
\end{equation}
which is just Dirac's eigenbra equation (\ref{braeqeneintro}). Thus,
Dirac's eigenbra equation acquires a precise meaning through
Eq.~(\ref{eeRHSb}), in the sense that it has to be understood as 
``right sandwiched'' with the wave functions $\varphi$ of $\mathbf \Phi$. 

By using definition~(\ref{eeRHSb}),
one can show that $\langle p|$, $\langle x|$ and $_{\rm l,r} \langle ^{\pm}E|$
are indeed left eigenvectors of $P$, $Q$ and $H$, respectively~\cite{FOCO}:
\begin{equation}
       \langle p|P=p\langle p| \, , \quad p\in \mathbb R \, ,
        \label{bpeP}
\end{equation}
\begin{equation}
      \langle x|Q=x\langle x| \, , \quad x \in \mathbb R \, ,
\end{equation}
\begin{equation}
       _{\rm l,r} \langle ^{\pm}E|H=
       E \hskip0.12cm  
         _{\rm l,r} \langle ^{\pm}E|    \, , \quad E\in [0,\infty )  \, .
        \label{bEeH}
\end{equation}

It is worthwhile noting that, in accordance with 
Dirac's formalism, there is a one-to-one correspondence between bras and 
kets~\cite{ONEONEROBERTS}; that is, given an observable $A$, to each element 
$a$ in the 
spectrum of $A$ there correspond a bra $\langle a|$ that is a left eigenvector
of $A$ and also a ket $|a\rangle$ that is a right eigenvector of $A$. The bra 
$\langle a|$ belongs to $\mathbf \Phi ^{\prime}$, whereas the ket $|a\rangle$ 
belongs to $\mathbf \Phi ^{\times}$.

\subsection{The Dirac basis expansions}

A crucial ingredient of Dirac's formalism is that the bras and kets of an
observable form a complete basis system, see Eqs.~(\ref{introDirbaexp}) and 
(\ref{introresiden}). When applied to $P$, $Q$ and $H$, 
Eq.~(\ref{introresiden}) yields
\begin{equation}
      \int_{-\infty}^{\infty}  \rmd p \, |p\rangle \langle p| = I \, ,
      \label{resonidentP}
\end{equation}
\begin{equation}
      \int_{-\infty}^{\infty}  \rmd x' \, |x'\rangle \langle x'| = I \, ,
        \label{resonidentQ}
\end{equation}
\begin{equation}
      \int_{0}^{\infty} \rmd E \, |E^{\pm}\rangle _{\rm l}\, 
                       _{\rm l}\langle ^{\pm}E| +
      \int_{0}^{\infty}\rmd E \, |E^{\pm}\rangle _{\rm r}\, 
                       _{\rm r}\langle ^{\pm}E| =  I \, ,
        \label{resonidentH}
\end{equation}
In the present subsection, we derive various Dirac basis expansions 
for the algebra of the 1D rectangular barrier potential. We will do so by
formally sandwiching Eqs.~(\ref{resonidentP})-(\ref{resonidentH}) in between
different vectors.

If we sandwich Eqs.~(\ref{resonidentP})-(\ref{resonidentH}) in between
$\langle x|$ and $\varphi$, we obtain
\begin{equation}
      \langle x|\varphi \rangle= \int_{-\infty}^{\infty}  \rmd p \, 
         \langle x|p\rangle \langle p|\varphi \rangle  \, ,
      \label{resonidentPxphi}
\end{equation}
\begin{equation}
      \langle x|\varphi \rangle = \int_{-\infty}^{\infty}  \rmd x' \, 
        \langle x|x'\rangle \langle x'|\varphi \rangle   \, ,
        \label{resonidentQxphi}
\end{equation}
\begin{equation}
     \langle x|\varphi \rangle = \int_{0}^{\infty}  \rmd E \, 
         \langle x|E^{\pm}\rangle _{\rm l} \, 
            _{\rm l}\langle ^{\pm}E|\varphi \rangle +
      \int_{0}^{\infty}  \rmd E \, 
           \langle x|E^{\pm}\rangle _{\rm r} \,  
           _{\rm r}\langle ^{\pm}E|\varphi \rangle \, .
        \label{resonidentHxphi}
\end{equation}
Equations~(\ref{resonidentPxphi})-(\ref{resonidentHxphi}) can be rigorously
proved by way of the RHS~\cite{FOCO}. In proving
these equations, we give meaning to 
Eqs.~(\ref{resonidentP})-(\ref{resonidentH}), which are just formal 
equations: Equations~(\ref{resonidentP})-(\ref{resonidentH}) have always to be
understood as part of a ``sandwich.'' Note
that Eqs.~(\ref{resonidentPxphi})-(\ref{resonidentHxphi})
are not valid for every element of the Hilbert space but only for those 
$\varphi$ that belong to $\Sw$, because the action 
of the bras and kets is well defined only on $\Sw$~\cite{EXTOHS}. Thus, the 
RHS, rather than just the Hilbert space, fully justifies the Dirac basis
expansions. Physically, the Dirac basis expansions provide the means to
visualize wave packet formation out of a continuous linear superposition
of bras and kets.

We can obtain similar expansions to 
Eqs.~(\ref{resonidentPxphi})-(\ref{resonidentHxphi}) by sandwiching
Eqs.~(\ref{resonidentP})-(\ref{resonidentH}) in between other vectors. For 
example, sandwiching Eq.~(\ref{resonidentQ}) in between $\langle p|$ and 
$\varphi$ yields~\cite{FOCO}
\begin{equation}
      \langle p|\varphi \rangle =
     \int_{-\infty}^{\infty}\rmd x \
        \langle p|x\rangle 
      \langle x|\varphi \rangle  \, ,
    \label{inveqDvaeipx}
\end{equation}
and sandwiching Eq.~(\ref{resonidentQ}) in between 
$_{\rm l,r}\langle ^{\pm}E|$ and $\varphi$ yields~\cite{FOCO}
\begin{equation}
      _{\rm l,r}\langle ^{\pm}E|\varphi \rangle =
     \int_{-\infty}^{\infty}\rmd x \
        _{\rm l,r}\langle ^{\pm}E|x\rangle 
      \langle x|\varphi \rangle  \, .
    \label{inveqDvaeix}
\end{equation}
It is worthwhile noting the parallel between the Dirac basis 
expansions and the Fourier expansions (\ref{resonidentPxphi}) and
(\ref{inveqDvaeipx})~\cite{FOCO}. This parallel will be used in 
Sec.~\ref{sec:phymean} to physically interpret the Dirac bras and kets.

We can also sandwich  Eqs.~(\ref{resonidentP})-(\ref{resonidentH}) in between
two elements $\psi$ and $\varphi$ of $\Sw$, and obtain~\cite{FOCO}
\begin{equation}
      (\varphi ,\psi ) = \int_{-\infty}^{\infty}\rmd p \, 
      \langle \varphi |p\rangle  \langle p|\psi \rangle  \, ,
       \label{spinofpbk}
\end{equation}
\begin{equation}
      (\varphi ,\psi ) = \int_{-\infty}^{\infty}\rmd x \, 
      \langle \varphi |x\rangle  \langle x|\psi \rangle  \, ,
       \label{spinofxbk}
\end{equation}
\begin{equation}
      (\varphi ,\psi ) = \int_0^{\infty}\rmd E\, 
      \langle \varphi |E^{\pm}\rangle_{\rm l}\, 
       _{\rm l}\langle ^{\pm}E|\psi \rangle +
       \int_0^{\infty}\rmd E\, 
      \langle \varphi |E^{\pm}\rangle_{\rm r}\, 
       _{\rm r}\langle ^{\pm}E|\psi \rangle  \, .
           \label{spinofEbk}   
\end{equation}
Equations~(\ref{spinofpbk})-(\ref{spinofEbk}) allow us to calculate the
overlap of two wave functions $\varphi$ and $\psi$ by way of the action 
of the bras and kets on those wave functions.

The last aspect of Dirac's formalism we need to implement is 
prescription~(\ref{introactionA}), which expresses the action of an observable
$A$ in terms of the action of its bras and kets. When applied to $P$, $Q$ and
$H$, prescription~(\ref{introactionA}) yields
\begin{equation}
      P = \int_{-\infty}^{\infty}\rmd p \, p |p\rangle  \langle p| \, ,
      \label{presAP}
\end{equation} 
\begin{equation}
      Q = \int_{-\infty}^{\infty}\rmd x \, x |x\rangle  \langle x| \, ,
      \label{presAQ}
\end{equation} 
\begin{equation}
      H = \int_{0}^{\infty}\rmd E \, E |E^{\pm}\rangle_{\rm l}\, 
       _{\rm l}\langle ^{\pm}E| +
        \int_{0}^{\infty}\rmd E \, E |E^{\pm}\rangle_{\rm r}\, 
       _{\rm r}\langle ^{\pm}E| \, .
     \label{presAH}
\end{equation}  
Needless to say, these equations are formal expressions that acquire meaning 
when properly sandwiched. For example, sandwiching them in between
$\langle x|$ and $\varphi$ yields~\cite{FOCO}
\begin{equation}
      \langle x|P \varphi \rangle =\int_{-\infty}^{\infty}\rmd p \, p
      \langle x |p\rangle  \langle p|\varphi \rangle \, , 
        \label{GMT2xP}
\end{equation}
\begin{equation}
        \langle x|Q \varphi \rangle = \int_{-\infty}^{\infty}\rmd x' \, x'
      \langle x |x'\rangle  \langle x'|\varphi \rangle \, ,
       \label{GMT2xQ}
\end{equation}
\begin{equation}
      \langle x|H \varphi \rangle = \int_0^{\infty} \rmd E \,
      E \langle x |E^{\pm}\rangle_{\rm l}\, 
       _{\rm l}\langle ^{\pm}E|\varphi \rangle +
       \int_0^{\infty}\rmd E\, E
      \langle x |E^{\pm}\rangle_{\rm r}\, 
       _{\rm r}\langle ^{\pm}E|\varphi \rangle  \, ,
        \label{GMT2xH}
\end{equation}
and sandwiching them in between two
elements $\varphi$ and $\psi$ of $\Sw$ yields~\cite{FOCO}
\begin{equation}
     (\varphi ,P \psi )=\int_{-\infty}^{\infty}\rmd p \, p
      \langle \varphi |p\rangle  \langle p|\psi \rangle \, , 
        \label{GMT2P}
\end{equation}
\begin{equation}
     (\varphi ,Q \psi )=\int_{-\infty}^{\infty}\rmd x \, x
      \langle \varphi |x\rangle  \langle x|\psi \rangle \, ,
       \label{GMT2Q}
\end{equation}
\begin{equation}
     (\varphi ,H \psi )= \int_0^{\infty} \rmd E \,
      E \langle \varphi |E^{\pm}\rangle_{\rm l}\, 
       _{\rm l}\langle ^{\pm}E|\psi \rangle +
       \int_0^{\infty}\rmd E\, E
      \langle \varphi |E^{\pm}\rangle_{\rm r}\, 
       _{\rm r}\langle ^{\pm}E|\psi \rangle  \, .
        \label{GMT2H}
\end{equation}
Note that, in particular, the operational definition of an 
observable---according to which an observable is simply an operator whose 
eigenvectors form a complete basis such that Eqs.~(\ref{introDirbaexp}), 
(\ref{introresiden}) and~(\ref{introactionA}) hold, see for example 
Ref.~\cite{COHEN}---acquires meaning within the~RHS.

The sandwiches we have made so far always involved at least a wave function
$\varphi$ of $\Sw$. When
the sandwiches do not involve elements of $\Sw$ at all, we obtain expressions
that are simply formal. These formal expressions are often useful though,
because they help us understand the meaning of concepts such as the delta 
normalization or the ``matrix elements'' of an operator. Let us start with the 
meaning of the delta normalization. When we sandwich Eq.~(\ref{resonidentQ}) 
in between $\langle p'|$ and $|p\rangle$, we get
\begin{equation}
      \int_{-\infty}^{\infty}  \rmd x \, 
         \langle p'|x\rangle \langle x|p\rangle = \langle p'|p \rangle \, .
        \label{resonidentQsppp}
\end{equation}
This equation is a formal expression that is to be understood in a
distributional sense, that is, both sides must appear smeared out by a smooth
function $\varphi (p) =\langle p|\varphi \rangle$ in an integral over $p$:
\begin{equation}
     \int_{-\infty}^{\infty}  \rmd p \, \varphi (p)
     \int_{-\infty}^{\infty}  \rmd x \, 
         \langle p'|x\rangle \langle x|p\rangle = 
     \int_{-\infty}^{\infty}  \rmd p \, \varphi (p) \langle p'|p \rangle \, .
        \label{smeresonidentQsppp}
\end{equation}
The left-hand side of Eq.~(\ref{smeresonidentQsppp}) can be written as
\begin{eqnarray}
    \int_{-\infty}^{\infty}  \rmd x \,  \langle p'|x\rangle
     \int_{-\infty}^{\infty}  \rmd p \, \varphi (p)
         \langle x|p\rangle  
    &=& 
        \int_{-\infty}^{\infty}  \rmd x \,  \langle p'|x\rangle
     \int_{-\infty}^{\infty} \rmd p \, 
          \langle x|p\rangle\langle p|\varphi \rangle \nonumber \\ 
    &=& 
     \int_{-\infty}^{\infty}  \rmd x \,  
         \langle p'|x\rangle \langle x|\varphi \rangle \nonumber \\
    &=& \varphi (p')
        \label{formdelnp}
\end{eqnarray}
Plugging Eq.~(\ref{formdelnp}) into Eq.~(\ref{smeresonidentQsppp}) leads to
\begin{equation}
     \int_{-\infty}^{\infty}  \rmd p \, \varphi (p) \langle p'|p\rangle =
         \varphi (p') \, .
           \label{vpapaovap}
\end{equation}
By recalling the definition of the delta function, we see that 
Eq.~(\ref{vpapaovap}) leads to
\begin{equation}
     \langle p'|p\rangle = \delta (p-p') \, ,
           \label{pppdeltappp}
\end{equation}
and to
\begin{equation}
     \int_{-\infty}^{\infty}\rmd x \, \langle p'|x\rangle \langle x|p\rangle = 
          \delta (p-p') \, .
           \label{pppdeltappp2}
\end{equation}
By using Eq.~(\ref{pxiscmplxp}), we can write Eq.~(\ref{pppdeltappp2}) in
a well-known form:
\begin{equation}
      \frac{1}{2\pi \hbar} \int_{-\infty}^{\infty}  
         \rmd x \, \rme^{\rmi(p-p')x/\hbar} = \delta (p-p') \, .
           \label{pppdeltappp3}
\end{equation}
This formal equation is interpreted by saying that the bras and kets of the 
momentum operator are delta normalized. That the energy bras and kets are 
also delta normalized can be seen in a similar, though slightly more involved 
way~\cite{FP02}:
\numparts
\begin{equation}
      _{\alpha}\langle ^{\pm}E'|E^{\pm} \rangle _{\beta} =
         \delta (E-E') \, \delta _{\alpha \beta} \, ,
              \label{deltanornebk1}
\end{equation}
\begin{equation}
      \int_{-\infty}^{\infty}  \rmd x \  _{\alpha}\langle ^{\pm}E'|x\rangle 
      \langle x|E^{\pm} \rangle _{\beta} =
         \delta (E-E') \, \delta _{\alpha \beta} \, ,
                \label{deltanornebk2}
\end{equation} 
\endnumparts
where $\alpha, \beta$ stand for the labels ${\rm l,r}$ that respectively 
denote left and right incidence. The derivation of 
expressions involving the Dirac delta function such as 
Eqs.~(\ref{pppdeltappp}), (\ref{pppdeltappp3}) 
or~(\ref{deltanornebk1})-(\ref{deltanornebk2}) shows
that these formal expressions must be understood in a distributional sense, 
that is, as kernels of integrals that include the wave functions $\varphi$ 
of $\Sw$, like in Eq.~(\ref{smeresonidentQsppp}). 

In a similar way, we can also understand the meaning of the 
``matrix elements'' of the observables in a particular representation, e.g.:
\begin{equation}
        \langle x|Q|x'\rangle = x' \, \delta (x-x') \, ,
         \label{meQxx}
\end{equation}
\begin{equation}
        \langle x|P|x'\rangle = -\rmi \hbar \frac{\rmd}{\rmd x} \ 
                       \delta (x-x') \, ,
        \label{mePxx}
\end{equation}
\begin{equation}
        \langle x|H|x'\rangle = 
    \left( -\frac{\hbar ^2}{2m}\frac{\rmd ^2}{\rmd x^2}+V(x) \right) 
            \delta (x-x')   \, .
     \label{meHxx}
\end{equation}
Equations~(\ref{meQxx})-(\ref{meHxx}) can be obtained by formally inserting
Eq.~(\ref{resonidentQ}) into respectively Eq.~(\ref{fdopx}), (\ref{fdopp})
and (\ref{fdoph}). 

It is illuminating to realize that the expressions~(\ref{meQxx})-(\ref{meHxx}) 
generalize the matrix representation of an observable $A$ in a 
finite-dimensional Hilbert space. If $a_1, \ldots ,a_N$ are the eigenvalues 
of $A$, then, in the basis $\{ |a_1 \rangle , \ldots , |a_N \rangle \}$, 
$A$ is represented as
\begin{equation}
      A\equiv \left( \begin{array}{cccc}
                           a_1 & 0 & \cdots & 0 \\
                           0&  a_2 & \cdots & 0 \\ 
                           \cdots & \cdots & \cdots & \cdots \\
                           0&  0 & \cdots & a_N 
                       \end{array}
                \right) \, ,
\end{equation}
which in Dirac's notation reads as
\begin{equation} 
      \langle a_i|A|a_j\rangle = a_i \delta _{ij} \, .
       \label{matreleA}
\end{equation}
Clearly, expressions~(\ref{meQxx})-(\ref{meHxx}) are the infinite-dimensional
extension of expression~(\ref{matreleA}).

\section{Physical meaning of the Dirac bras and kets}
\label{sec:phymean}

The bras and kets associated with eigenvalues in the continuous spectrum are 
not normalizable. Hence, the standard probabilistic interpretation does not 
apply to them straightforwardly. In this section, we 
are going to generalize the probabilistic interpretation of normalizable 
states to the non-normalizable bras and kets. As well, in order to gain
further insight into the physical meaning of bras and kets, we shall present
the analogy between classical plane waves and the bras and kets.

In Quantum Mechanics, the scalar product of the Hilbert space is employed to 
calculate probability amplitudes. In our example, the Hilbert space is
$L^2$, and the corresponding scalar product is given
by Eq.~(\ref{scapro}). That an eigenvalue of an observable $A$ lies in the 
discrete or in the continuous part of the spectrum is determined by this 
scalar product. An eigenvalue $a_n$ belongs to the discrete part of the spectrum when its 
corresponding eigenfunction $f_n(x)\equiv \langle x|a_n\rangle$ is square 
normalizable:
\begin{equation}
      (f_n,f_n)=\int_{-\infty}^{\infty}\rmd x \, |f_n(x)|^2 <\infty  \, .
          \label{pspn}
\end{equation}
An eigenvalue $a$ belongs to the continuous part of the spectrum when its 
corresponding eigenfunction $f_a(x)\equiv \langle x|a\rangle$ is {\it not} 
square normalizable:
\begin{equation}
      (f_a,f_a)=\int_{-\infty}^{\infty}\rmd x \, |f_a(x)|^2 =\infty  \, .
          \label{pspE}
\end{equation}
In the latter case, one has to use the
theory of distributions to ``normalize'' these states, e.g., delta function
normalization:
\begin{equation}
      (f_a,f_{a'})=\int_{-\infty}^{\infty}\rmd x \, 
              \overline{f_a(x)} f_{a'}(x) =\delta (a-a') \, .
          \label{pspEdn}
\end{equation}
This Dirac delta normalization generalizes the Kronecker delta normalization
of ``discrete'' states:
\begin{equation}
      (f_n,f_{n'})=\int_{-\infty}^{\infty}\rmd x \, 
              \overline{f_n(x)} f_{n'}(x) =\delta _{nn'}   \, .
\end{equation}
Because they are square integrable, the ``discrete'' eigenvectors 
$f_n(x)\equiv \langle x|a_n\rangle$ can be
interpreted in the usual way as probability amplitudes. But because they are 
{\it not} square integrable, the ``continuous'' eigenvectors 
$f_a(x)\equiv \langle x|a\rangle$ must be interpreted as ``kernels'' 
of probability amplitudes, in the sense that when we multiply 
$\langle x|a\rangle$ by $\langle \varphi |x \rangle$ and then integrate, we
obtain the density of probability amplitude $\langle \varphi |a\rangle$:
\begin{equation}
   \langle \varphi |a\rangle = \int_{-\infty}^{\infty}\rmd x \, 
    \langle \varphi |x \rangle  \langle x|a\rangle \, .
\end{equation}
Thus, in particular, $\langle x|p\rangle$, $\langle x|x'\rangle$ and 
$\langle x|E^{\pm}\rangle _{\rm l,r}$ represent ``kernels''
of probability amplitudes.

Another way to interpret the bras and kets
is in analogy to the plane waves of classical optics and classical 
electromagnetism. Plane waves $\rme^{\rmi kx}$ represent monochromatic light 
pulses of wave number $k$ and frequency (in vacuum) $w=kc$. Monochromatic light
pulses are impossible to prepare experimentally; all that can be prepared
are light pulses $\varphi (k)$ that have some wave-number spread. The 
corresponding pulse in the position representation, $\varphi (x)$, can be 
``Fourier decomposed'' in terms of the monochromatic plane waves as
\numparts
\begin{equation}
  \varphi (x) = \frac{1}{\sqrt{2\pi }} \int \rmd k \, 
                    \rme^{\rmi kx}\varphi (k) \, ,
       \label{fourinnet}
\end{equation}
which in Dirac's notation becomes
\begin{equation}
      \langle x|\varphi \rangle = \int \rmd k \, 
     \langle x|k \rangle \langle k|\varphi \rangle   \, .
\end{equation}
\endnumparts
Thus, physically preparable pulses can be expanded in a Fourier
integral by the unpreparable plane waves, the weights of the
expansion being $\varphi (k)$. When $\varphi (k)$ is highly peaked around
a particular wave number $k_0$, then the pulse can in general be 
represented for all practical purposes by a monochromatic plane wave 
$\rme ^{\rmi k_0x}$. Also, in finding out how a light pulse behaves under 
given conditions (e.g., reflection and refraction at a plane interface between 
two different media), we only have to find out how plane waves behave 
and, after that, by means of the Fourier expansion~(\ref{fourinnet}), we 
know how the light pulse $\varphi (x)$ behaves. Because obtaining the 
behavior of plane waves is somewhat easy, it is advantageous to use them to 
obtain the behavior of the whole pulse~\cite{USEFULNESSPW}.

The quantum mechanical bras and kets can be interpreted in analogy 
to the classical plane waves. The eigenfunction 
$\langle x|p\rangle = \rme ^{\rmi px/\hbar} / \sqrt{2\pi \hbar}$ represents a 
particle of sharp momentum $p$; the eigenfunction 
$\langle x|x'\rangle = \delta (x-x')$ represents a particle sharply localized
at $x'$; the monoenergetic eigenfunction 
$\langle x|E^{\pm}\rangle _{\rm l,r}$ represents a particle with well-defined 
energy $E$ (and with additional boundary conditions determined by the
labels $\pm$ and ${\rm l,r}$). In complete analogy to the Fourier
expansion of a light pulse by classical plane waves,
Eq.~(\ref{fourinnet}),
the eigenfunctions $\langle x|p\rangle$, $\langle x|x'\rangle$ and
$\langle x|E^{\pm}\rangle _{\rm l,r}$ expand a wave function $\varphi$,
see Eqs.~(\ref{resonidentPxphi})-(\ref{resonidentHxphi}). When the wave 
packet $\varphi (p)$ is highly peaked around a particular 
momentum $p_0$, then in general the approximation 
$\varphi (x) \sim \rme ^{\rmi p_0x/\hbar} / \sqrt{2\pi \hbar}$ holds for all
practical purposes; when the wave packet $\varphi (x)$ is highly peaked around 
a particular position $x_0$, then in general the approximation 
$\varphi (x) \sim \delta (x-x_0)$ holds for all practical purposes; and when 
$\varphi (E)$ is highly peaked around a particular energy $E_0$, 
then in general the approximation 
$\varphi (x) \sim \langle x|E_0^{\pm}\rangle _{\rm l,r}$ holds for all
practical purposes (up to the boundary conditions determined by the
labels $\pm$ and ${\rm l,r}$). Thus, although in principle 
$\langle x|p\rangle$, $\langle x|x'\rangle$ and 
$\langle x|E^{\pm}\rangle _{\rm l,r}$ are impossible to prepare, in many
practical situations they can give good approximations when the wave packet
is well peaked around some particular values $p_0$, $x_0$, $E_0$ of the
momentum, position and energy. Also, in finding out how a wave function 
behaves under given conditions (e.g., reflection and transmission off a 
potential barrier), all we have to find out is how the bras and kets behave 
and, after that, by means of the Dirac basis expansions, we know how the 
wave function $\varphi (x)$ behaves. Because obtaining the behavior of 
the bras and kets is somewhat easy, it is advantageous to use them to obtain
the behavior of the whole wave function~\cite{USEFULNESSBK}.

From the above discussion, it should be clear that
there is a close analogy between classical Fourier methods and
Dirac's formalism. In fact, one can say that Dirac's formalism is the extension
of Fourier methods to Quantum Mechanics: Classical monochromatic plane waves 
correspond to the Dirac bras and kets; the light pulses correspond to the
wave functions $\varphi$; the classical Fourier expansion corresponds to
the Dirac basis expansions; the classical Fourier expansion provides the 
means to form light pulses out of a continuous linear superposition of 
monochromatic plane waves, and the Dirac basis expansions provide the 
means to form wave functions out of a continuous linear superposition of
bras and kets; the classical uncertainty principle of Fourier
Optics corresponds to the quantum uncertainty generated by the 
non-commutativity of two observables~\cite{DEBROGLIE}. However, although this 
analogy is very close from a formal point of view, there is a crucial 
difference from a conceptual point of view. To wit, whereas in the classical 
domain the solutions of the wave equations represent a physical wave, in 
Quantum Mechanics the solutions of the equations do {\it not} represent a 
physical object, but rather a probability amplitude---In Quantum Mechanics what
is ``waving'' is probability.

\section{Further considerations}
\label{sec:gener}

In Quantum Mechanics, the main objective is to obtain the probability of 
measuring an observable $A$ in a state $\varphi$. Within the Hilbert space
setting, such probability can be obtained by means of the spectral
measures ${\sf E}_a$ of $A$ (see, for example, Ref.~\cite{GALINDO}). These 
spectral measures satisfy
\begin{equation}
        I = \int_{{\rm Sp}(A)} \rmd {\sf E}_a \, 
\end{equation}
and 
\begin{equation}
        A = \int_{{\rm Sp}(A)} a \,  \rmd {\sf E}_a \, 
\end{equation}
Comparison of these equations with Eqs.~(\ref{introresiden}) and 
(\ref{introactionA}) yields
\begin{equation}
       \rmd {\sf E}_a = |a\rangle \langle a| \, \rmd a \, .
           \label{factorizationofE}
\end{equation}
Thus, the RHS is able to ``factor out'' the Hilbert space spectral measures in 
terms of the bras and kets~\cite{FSHORT}. For the position, momentum and 
energy observables, Eq.~(\ref{factorizationofE}) reads as
\begin{equation}
       \rmd {\sf E}_x = |x\rangle \langle x| \,  \rmd x \, ,
\end{equation}
\begin{equation}
       \rmd {\sf E}_p = |p\rangle \langle p| \, \rmd p \, ,
\end{equation}
\begin{equation}
       \rmd {\sf E}_E = |E^{\pm} \rangle _{\rm l} \,  
                             _{\rm l}\langle ^{\pm} E| \,  \rmd E +
        |E^{\pm} \rangle_{\rm r} \,  _{\rm r}  \langle ^{\pm} E| \, \rmd E   
                     \, .
       \label{spmesH}
\end{equation}
Although the spectral measures $\rmd {\sf E}_a$ associated with a given
self-adjoint operator $A$ are unique, the factorization
in terms of bras and kets is not. For example, as we can see from
Eq.~(\ref{spmesH}), the spectral measures of our Hamiltonian can be written in 
terms of the basis $\{ |E^{+} \rangle _{\rm l,r} \}$ or the basis 
$\{ |E^{-} \rangle _{\rm l,r} \}$. From a physical point of view, those
two basis are very different. As we saw in Sec.~\ref{sec:e1dsbp}, the basis 
$\{ |E^{+} \rangle _{\rm l,r} \}$
represents the initial condition of an incoming particle, whereas the basis 
$\{ |E^{-} \rangle _{\rm l,r} \}$ represents the final condition of an 
outgoing particle. However, the spectral measures of the Hilbert space are 
insensitive to such difference, in contrast to the RHS, which can differentiate
both cases. Therefore, when computing probability amplitudes, the RHS gives 
more precise information on how those probabilities are physically produced 
than the Hilbert space.

In this paper, we have restricted our discussion to the simple, 
straightforward algebra
of the 1D rectangular barrier. But, what about more complicated potentials? In
general, the situation is not as easy. First, the theory of rigged Hilbert 
spaces as constructed by Gelfand and collaborators is based on the assumption
that the space $\mathbf \Phi$ has a property 
called {\it nuclearity}~\cite{GELFAND,MAURIN}. However, it is not clear 
that one can always find a nuclear space $\mathbf \Phi$ that remains invariant 
under the action of the observables. Nevertheless, Roberts has shown that
such $\mathbf \Phi$ exists when the potential is infinitely often 
differentiable except for a closed set of zero Lebesgue 
measure~\cite{ROBERTS}. Second, the problem of constructing the 
RHS becomes more involved when the observable $A$ is not 
cyclic~\cite{GELFAND}. And third, solving the eigenvalue
equation of an arbitrary self-adjoint operator is rarely as easy as in our 
example.

\section{Summary and conclusions}
\label{sec:conclusions}

We have used the 1D rectangular barrier model to see that, when the spectra of
the observables have a continuous part, the natural setting for Quantum 
Mechanics is the rigged Hilbert space rather than just the Hilbert space. In
particular, Dirac's bra-ket formalism is fully implemented by the rigged
Hilbert space rather than just by the Hilbert space. 

We have explained the physical and mathematical meanings of each of
the ingredients that form the rigged Hilbert space. Physically, the space 
$\mathbf \Phi \equiv \Sw$ is interpreted as the space of wave 
functions, since its elements can be 
associated well-defined, finite physical quantities, and algebraic operations
such as commutation relations are well defined on 
$\mathbf \Phi$. Mathematically, $\mathbf \Phi$ is the space of 
test functions. The spaces ${\mathbf \Phi}^{\prime} \equiv \Swp$ and 
${\mathbf \Phi}^{\times} \equiv \Swt$ contain respectively the bras and kets 
associated with the eigenvalues that lie in the continuous 
spectrum. Physically, the bras and kets are interpreted as ``kernels'' of 
probability amplitudes. Mathematically, the bras and kets are 
distributions. The following table summarizes the meanings of each space:
\begin{center}
\begin{tabular}{|c|l|l|}
      \hline
     {\sc Space} & \, {\sc Physical Meaning} & \, {\sc Mathematical Meaning}
                                                                    \, \\
      \hline 
      \hline
     ${\mathbf \Phi}$ & \, Space of wave functions $\varphi$ \,  & 
             \, Space of test functions $\varphi$ \,  \\
     ${\cal H}$ & \, Probability amplitudes & \, Hilbert space \\
     $\, \, {\mathbf \Phi}^{\times}$ & \, Space of kets $|a\rangle$  & 
                                                        \, Antidual space \\
     $\, {\mathbf \Phi}^{\prime}$ & \, Space of bras $\langle a|$ & 
                                                         \, Dual space  \\
     \hline
\end{tabular}
\end{center}

We have seen that, from a physical point of view, the rigged Hilbert 
space does not entail an extension of Quantum Mechanics, whereas, from a 
mathematical point of view, the rigged Hilbert space is an extension of
the Hilbert space. Mathematically, the rigged Hilbert space 
arises when we equip the Hilbert space with distribution theory. Such 
equipment enables us to cope with singular objects such as bras and kets.

We have also seen that formal expressions involving bras and kets
must be understood as ``sandwiched'' by wave functions $\varphi$. Such
``sandwiching'' by $\varphi$'s is what controls the singular behavior of
bras and kets. This is why mathematically the sandwiching by $\varphi$'s is 
so important and must always be implicitly assumed. In practice, we can 
freely apply the formal manipulations of Dirac's formalism with confidence, 
since such formal manipulations are justified by the rigged Hilbert space.

We hope that this paper can serve as a pedagogical, enticing introduction to 
the rigged Hilbert space.

\ack

Research supported by the Basque Government through reintegration
fellowship No.~BCI03.96, and by the University of the Basque Country
through research project No.~9/UPV00039.310-15968/2004.

\newpage

\newpage

\begin{thebibliography}{99}
\bibitem{CITEMAURIN} In Ref.~\cite{MAURIN}, page~7, Maurin states that
``{\it It seems to us that this is the formulation 
which was anticipated by Dirac in his classic 
monograph.''}
\bibitem{DENSE} A subspace $S$ of $\cal H$ is dense in $\cal H$
if we can approximate any element of $\cal H$ by an element of $S$
as well as we wish. Thus, for any $f$ of $\cal H$ and for any small
$\epsilon >0$, we can find a $\varphi$ in $S$ such that 
$\| f-\varphi \| < \epsilon$. In physical terms, this inequality means that
we can replace $f$ by $\varphi$ within an accuracy $\epsilon$.
\bibitem{FUNCTIONAL} A function $F: {\mathbf \Phi} \to {\mathbb C}$ is called
a linear [respectively antilinear] functional over $\mathbf \Phi$ if for any 
complex numbers $\alpha , \beta$ 
and for any $\varphi , \psi \in \mathbf \Phi$, it holds that
$F(\alpha \varphi +\beta \psi)=\alpha F(\varphi) +\beta F(\psi )$ 
[respectively 
$F(\alpha \varphi +\beta \psi)=\alpha ^* F(\varphi) +\beta ^*F(\psi )$].
\bibitem{HSDEF} Strictly speaking, a Hilbert space possesses additional
properties (e.g., it must be complete with respect to the topology induced by
the scalar product). For a more technical definition of the Hilbert space,
see for example Ref.~\cite{DIS}.
\bibitem{UNB} An operator $A$ is bounded if there is some finite $K$ such
that $\| Af \| <K \|f \|$ for all $f\in \cal H$, where $\| \  \|$ denotes 
the Hilbert space
norm. When such $K$ does not exist, $A$ is said to be unbounded. For a 
detailed account of the properties of bounded and unbounded operators, see for
example Ref.~\cite{DIS}.
\bibitem{RS84} The mathematical reason why quantum mechanical unbounded
operators cannot be defined on all the vectors of the Hilbert space can be
found, for example, in Ref.~\cite{RS}, page~84.
\bibitem{INFENER} If we nevertheless insisted in for example calculating the 
expectation value~(\ref{exintrodispP}) for elements of $\cal H$ that are not 
in ${\cal D}(A)$, we would obtain an unphysical infinity value. For instance, 
if $A$ represents an unbounded Hamiltonian $H$, then the expectation 
value~(\ref{exintrodispP}) would be infinite for those $\varphi$ of $\cal H$ 
that lie outside of ${\cal D}(H)$. Because they have infinite 
energy, those states do not represent physically preparable wave packets.
\bibitem{SNHS} If they were in the Hilbert space, $|a\rangle$ and $\langle a|$
would be square integrable, and $a$ would belong to the discrete spectrum.
\bibitem{RS274} It is well known that Heisenberg's commutation relation
necessarily implies that either $P$ or $Q$ is unbounded. See, for example, 
Ref.~\cite{RS}, page~274.
\bibitem{ZERODERIV} The reason why the derivatives of $\varphi (x)$ must 
vanish at $x=a,b$ is that we want to be able to apply the Hamiltonian $H$
as many times as we wish. Since repeated applications of $H$ to $\varphi (x)$
involve the derivatives of $V(x)\varphi (x)$, and since $V(x)$ is 
discontinuous at $x=a,b$, the function $V(x)\varphi (x)$ is infinitely 
differentiable at $x=a,b$ only when the derivatives of $\varphi (x)$ 
vanish at $x=a,b$. For more details, see Ref.~\cite{ROBERTS}. The vanishing 
of the derivatives of $\varphi (x)$ at $x=a,b$ must be viewed as a 
mathematical consequence of the unphysical sharpness of the discontinuities of
the potential, rather than as a physical consequence of Quantum 
Mechanics. Note 
also that in standard numerical simulations, for example, Gaussian wave 
packets impinging on a rectangular barrier, one never sees that the wave
packet vanishes at $x=a,b$. This is due to the fact that on a Gaussian wave 
packet, the Hamiltonian~(\ref{fdoph}) can only be applied once.
\bibitem{CRVANISHES} We note that, when acting on elements $\varphi$ of
$\Sw$, the commutator $[H,P] =\rmi \hbar \frac{\partial V}{\partial x}$
reduces to $[H,P]=0$, due to the vanishing of the derivatives of $\varphi$ 
at $x=a,b$.
\bibitem{ONEONEROBERTS} We recall that some authors have erroneously claimed 
that ``there are more kets than bras''~\cite{ROBERTS}, and that therefore such
one-to-one correspondence between bras and kets does not hold.
\bibitem{EXTOHS} We can nevertheless  extend 
Eqs.~(\ref{resonidentPxphi}) and (\ref{resonidentHxphi}) to the 
whole Hilbert space $L^2$ by a limiting procedure, 
although the 
resulting expansions do not involve the Dirac bras and kets any more, but
simply the eigenfunctions of the differential operators.
\bibitem{USEFULNESSPW} This is one of the major reasons why plane waves 
are so useful in practical calculations.
\bibitem{USEFULNESSBK} This is one of the major reasons why bras and kets
are so useful in practical calculations.
\bibitem{DEBROGLIE} There are many other links between the classical and the
quantum worlds, such as for example the de Broglie relation $p=\hbar k$, which
entails a formal identity between the classical $\rme ^{\rmi kx}$ and
the quantum $\rme ^{\rmi px/\hbar}$ plane waves.
\bibitem{FSHORT} We recall that the direct integral decomposition of the 
Hilbert space falls short of such factorization, see Ref.~\cite{ANTOINE}.
\end{thebibliography}
\end{document}